\documentclass{aa}  

\usepackage{txfonts}
\usepackage{verbatim}
\usepackage{natbib}
\usepackage{array}
\usepackage{threeparttable}
\usepackage{placeins}
\AddToHook{begindocument/before}{\usepackage{hyperref}}
\usepackage{orcidlink}

\def\kms{km\,s$^{-1}$}

\def\msun{M$_{\odot}$}

\def\rsol{R$_{\odot}$}

\def\s{$\sigma$}

\def\vsini{$v \sin i$}

\def\s{$\sigma$}

\def\teff{T$_\mathrm{eff}$}
\def\logg{$\log g$}
\def\heabun{$Y_\mathrm{He}$}
\def\cabun{$\epsilon_\mathrm{C}$}
\def\nabun{$\epsilon_\mathrm{N}$}
\def\oabun{$\epsilon_\mathrm{O}$}
\def\siabun{$\epsilon_\mathrm{Si}$}
\def\vmac{$v_\mathrm{macro}$}

\def\lam{$\lambda$}

\def\halpha{H\,$\alpha$}

\def\hea{He\,{\sc i}}
\def\heb{He\,{\sc ii}}

\def\cc{C\,{\sc iii}}
\def\cd{C\,{\sc iv}}

\def\nc{N\,{\sc iii}}
\def\nd{N\,{\sc iv}}
\def\ne{N\,{\sc v}}

\def\ob{O\,{\sc ii}}
\def\oc{O\,{\sc iii}}

\def\fw{\textsc{fastwind}}
\def\cmfgen{\textsc{cmfgen}}
\def\powr{\textsc{PoWR}}

\def\specfann{\textsc{SpecFANN}}

\def\pyGA{\textsc{pyGA}}
\def\emcee{\textsc{emcee}}

\def\scipy{\textsc{scipy}}
\def\ultranest{\textsc{ultranest}}

\begin{document}

   \title{\specfann: Spectral Fitting via Artificial Neural Networks}
   \subtitle{I. A deep learning based \fw\ emulator and fitting suite}
    \titlerunning{SpecFANN: FASTWIND emulator and fitting suite}
    \authorrunning{M. Abdul-Masih et al.}

   \author{
            M. Abdul-Masih \inst{\ref{inst:iac}, \ref{inst:ull}} \orcidlink{0000-0001-6566-7568}
            \and C. Hawcroft \inst{\ref{inst:stsci}} 
            \and T. Lechien \inst{\ref{inst:mpa}}
            \and S. Simon-Diaz \inst{\ref{inst:iac}, \ref{inst:ull}}
            \and H. Sana \inst{\ref{inst:kul}}
            \and K. Deshmukh \inst{\ref{inst:kul}}
            \and J. Vrancken \inst{\ref{inst:kul}}
            \and J.\ I.\ Villase\~{n}or\inst{\ref{inst:mpia}}
            \and J. M\"uller-Horn\inst{\ref{inst:mpia}}
            \and A. J. Kalita \inst{\ref{inst:NU}}
            \and B. Ludwig \inst{\ref{inst:kul}}
            \and J. Bodensteiner \inst{\ref{inst:api}}
            \and D. M. Bowman \inst{\ref{inst:NU}, \ref{inst:kul}} \orcidlink{0000-0001-7402-3852}
            \and A. Escorza \inst{\ref{inst:iac}, \ref{inst:ull}}
            \and G. Holgado \inst{\ref{inst:iac}, \ref{inst:ull}}
            \and J. Puls \inst{\ref{inst:lmu}}
            \and A. de Vicente \inst{\ref{inst:iac}, \ref{inst:ull}}
          }

   \institute{
   {Instituto de Astrofísica de Canarias, C. Vía Láctea, s/n, 38205 La Laguna, Santa Cruz de Tenerife, Spain \label{inst:iac}}\\ \email{mabdul@iac.es}
    \and
    {Universidad de La Laguna, Departamento de Astrofísica, Av. Astrofísico Francisco Sánchez s/n, 38206 La Laguna, Tenerife, Spain\label{inst:ull}}
    \and
    {Space Telescope Science Institute, 3700 San Martin Drive, Baltimore, MD 21218, USA \label{inst:stsci}}
    \and 
    {Max Planck Institute for Astrophysics, Karl-Schwarzschild-Strasse 1, 85748 Garching, Germany\label{inst:mpa}}
    \and
    {Institute of Astronomy, KU Leuven, Celestijnenlaan 200D, 3001 Leuven, Belgium\label{inst:kul}}
    \and
    {Max-Planck-Institut f\"{u}r Astronomie, K\"{o}nigstuhl 17, D-69117 Heidelberg, Germany\label{inst:mpia}}
    \and
    {{School of Mathematics, Statistics and Physics, Newcastle University, Newcastle upon Tyne, NE1 7RU, United Kingdom} \label{inst:NU}}
    \and 
    {{Anton Pannekoek Institute for Astronomy, University of Amsterdam, Science Park 904, 1098 XH Amsterdam, the Netherlands} \label{inst:api}}
    \and 
    {LMU München, Universitätssternwarte, Scheinerstr. 1, 81679 München, Germany \label{inst:lmu}}
    }
   \date{}

 
  \abstract
   {The importance of massive stars cannot be overstated: they are powerful probes of the early universe, they play a vital role in the chemical and mechanical evolution of their host environments and their end products allow us to study the most extreme physics in the universe. Obtaining accurate stellar and surface parameters for large samples of massive stars is vital to our understanding of how they evolve, and how their births, lives and deaths affect their surroundings. With the large volume of data expected from the next generation of spectroscopic surveys, it is likely that the computational cost of our current analysis methods (especially the calculation of model stellar atmospheres and synthetic spectra) will prove to be the most important bottleneck impeding our progress.}
   {In order to overcome these limitations and dramatically decrease computing times, we aim to develop a robust emulator for the \fw\ radiative transfer and spectral synthesis code. Additionally, we aim to explore alternative fitting methods that have not been feasible up to this point due to computational costs.}
   {We calculate a large set ($\sim$ 50\,000 models in total) of \fw\ synthetic spectra of OB-type stars, varying temperature, surface gravity helium mass fractions and CNOSi abundances, and we train a collection of neural networks to emulate these models with separate networks for each line.  We also develop the open-source python package \specfann, which provides users with a suite of fitting methods that can be used with these or other user-generated neural networks.}
   {The majority of the trained neural networks reach average accuracies of better than $\sim$0.01-0.1\% for photospheric lines and better than $\sim$0.1-1\% for wind lines. We find that \specfann\ is able to obtain robust and accurate stellar parameters that are consistent with the literature for a sample of 52 early-type stars.  Using \specfann\ we find that we can achieve the same fit in $\sim1/360\,000$ of the time when compared to alternative techniques that rely on `on-the-fly' \fw\ computations.}
   {We have demonstrated that neural networks offer a viable path forward to address the computational limitations of our current atmosphere analysis and stellar parameter determination methods for hot stars.}

   \keywords{Stars: massive -- 
                Methods: data analysis --
                Stars: atmospheres --
                Stars: fundamental parameters
                }

   \maketitle
   \nolinenumbers
%

\section{Introduction}\label{sec:intro}
   
While less numerous than their low-mass counterparts, massive stars exhibit extreme luminosities, allowing them to be studied from much farther distances, thus making them powerful probes of the distant and early universe \citep{haiman1997}.  Massive stars drive the mechanical and chemical evolution of their host environments through stellar winds and feedback, so understanding how these systems evolve, how long they live, and how they will die is essential to our understanding of the universe \citep{Bresolin2008, Hopkins2014}.  While great strides have been made in our theoretical understanding of massive star evolution, several important uncertainties remain, including internal mixing efficiencies \citep[e.g., ][]{Hunter2008, Schootemeijer2019, jin2024}, the outcomes and physics of binary interactions \citep[see review by ][]{Marchant2024} and many more.  Large sets of robust observational constraints are crucial to refine our models.

\begin{figure*}
    \centering
    \includegraphics[width=0.33\textwidth]{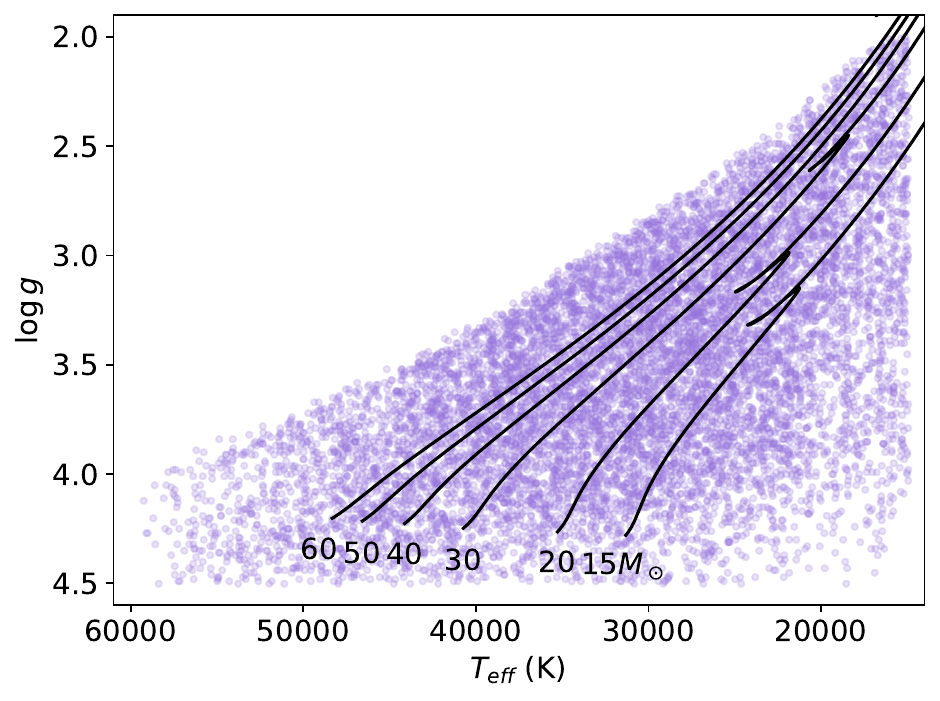}
    \includegraphics[width=0.33\textwidth]{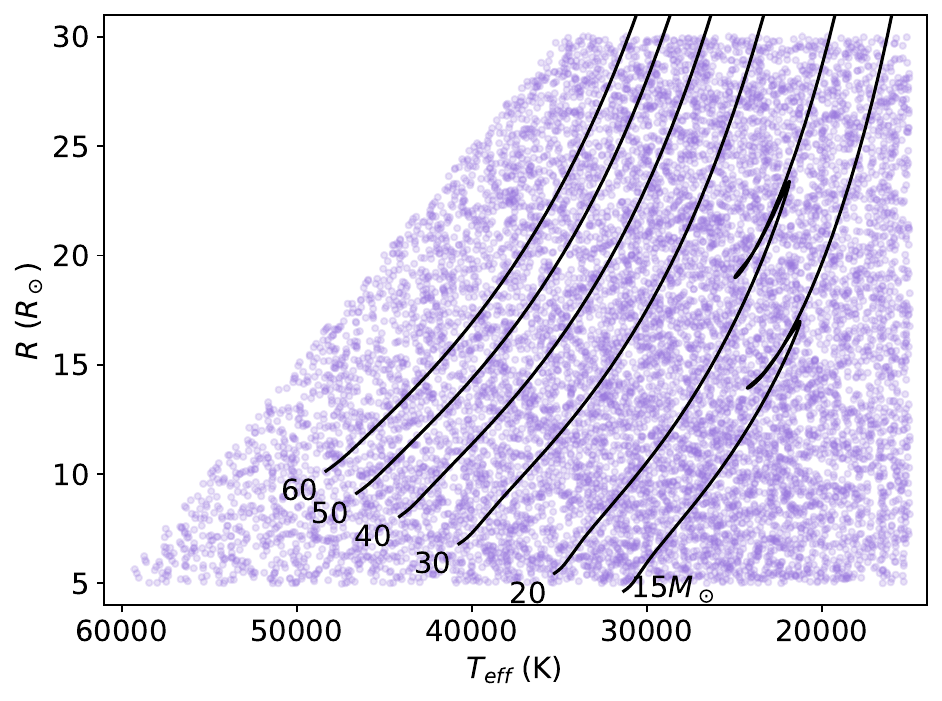}
    \includegraphics[width=0.33\textwidth]{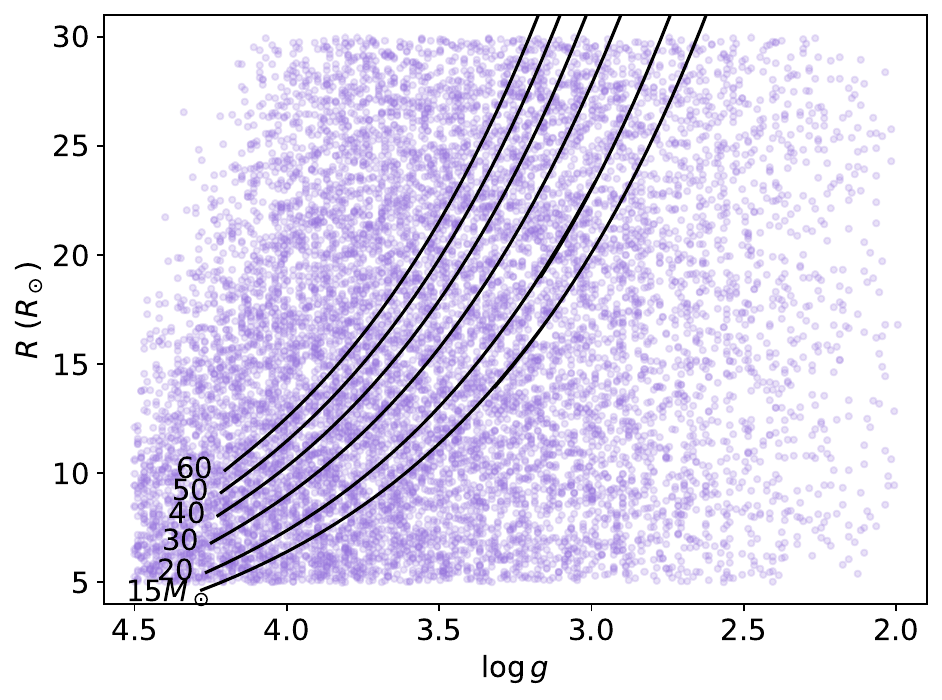}
    \caption{Parameter space covered by the \fw\ model set in surface gravity versus temperature (left), radius versus temperature (middle) and radius versus surface gravity (right).  Non-rotating evolutionary tracks from \citet{Brott2011b} are over-plotted in black for reference, with the mass of each track indicated in solar masses.}
    \label{fig: param_space}
\end{figure*}

Accurately measuring the stellar properties and surface abundances is fundamental to our understanding of massive star evolution as a whole.  Due to their high temperatures while on the main sequence, however, the assumption of local thermodynamic equilibrium (LTE), often used in spectral synthesis models of low-mass stars, is not valid in the massive star regime.  Instead, non-LTE radiative transfer codes are necessary to properly account for the atmospheric conditions and reproduce the observed spectral line profiles \citep{hubeny2014}. Several such codes, appropriate for the modelling of massive stars, are currently being actively developed, including but not limited to \fw\ \citep{Santolaya-Rey1997, Puls2005, RiveroGonzalez2012, Carneiro2016, Sundqvist2018a}, \cmfgen\ \citep{Hillier1998}, and \powr\ \citep{Grafener2002, Hamann2003, Sander2015}. However, the increased accuracy that these codes provide comes at the expense of computation time: calculating a single model can take anywhere from $\sim$1 hour to $\sim$1 day.  This severely limits the approaches we can reasonably use to fit these synthetic models to observed data. Considering the increased data flow each year from current and future multi-object spectrographs like WEAVE \citep{Dalton2014}, 4MOST \citep{deJong2019}, MOONS \citep{Cirasuolo2020}, MOSAIC \citep{Evans2015}, and related surveys like BLOeM \citep{shenar2024}, SDSS-V \citep{SDSSCollaboration2025, Kollmeier2026}, without a significant increase in speed, our progress in understanding massive stars may be bottlenecked by our ability to fit them with appropriate non-LTE radiative transfer codes.

Machine learning techniques offer a promising path forward \citep[see e.g., ][]{Urbaneja2026}. If adequately trained, deep neural networks offer efficient and accurate mapping between sets of input parameters and output values.  By applying such networks to non-LTE radiative transfer codes like those mentioned above, one can create a highly accurate emulator that can produce synthetic spectra in a fraction of the time. Not only can these networks allow us to handle the large datasets expected on the near horizon, they may also allow us to explore new fitting techniques that were not reasonably possible before.  

To this end, in this paper, we present a proof-of-concept neural network-based spectral synthesis emulator and the accompanying \specfann\  (Spectral Fitting via Artificial Neural Networks) code. This paper is organized as follows.  Section \ref{sec:methods} details the calculation of the synthetic spectra data set, the set-up of the neural networks, and the training procedure.  In Sect. \ref{sec:results}, we discuss the results of the training, and we validate the performance of the networks.  Section \ref{sec:specfan}\ introduces the \specfann\ code and details the included features, and Sect. \ref{sec:application}\ demonstrates an example application of the code.  In Sect. \ref{sec:discussion}, we discuss how \specfann\ compares to traditional fitting techniques, both in terms of accuracy and computation time, and we discuss how this method can be further improved.  Finally, in Sect. \ref{sec:conclusion}, we summarize and discuss future plans.

\section{Methods} \label{sec:methods}
With the goal of creating a neural network-based emulator for massive star spectral synthesis in mind, we choose to use the \linebreak \fw\ code (v10.6.5) to calculate the synthetic models needed to train the network.  \fw\ is a 1D non-LTE radiative transfer code that is suitable for the analysis of OBA stars \citep{Santolaya-Rey1997, Puls2005, RiveroGonzalez2012, Carneiro2016, Sundqvist2018a}.  Given a set of stellar parameters and surface abundances, \fw\ calculates the model atmosphere, wind structure, and corresponding spectral line profiles for a provided line list. Unlike other available non-LTE codes, \fw\ provides individual line profiles as opposed to fully integrated spectral windows, allowing for tailored networks for each line if needed.

\subsection{Calculation of FASTWIND models}

In order to cover as much of the massive star ($M > 8 M_{\odot}$) main sequence as possible, we only draw samples from certain areas of the full parameter space that \fw\ can cover.  Namely, our tailored parameter space encompasses temperatures between $T_\mathrm{eff} = $ 15~kK and 60~kK, surface gravities between $\log g = $ 2.0 and 4.5, radii between $R = $ 5~\rsol\ and 30~\rsol, helium abundances between $Y_\mathrm{He} = $ 0.06 and 0.30, and abundances of carbon, nitrogen, oxygen and silicon between $\epsilon_\mathrm{CNOSi} = $ 6.0 and 9.0. Note that we follow the abundance conventions used in \fw, namely that the helium abundance $Y_\mathrm{He}$ is defined as:

\begin{equation}
    Y_\mathrm{He} = \frac{N_\mathrm{He}}{N_\mathrm{H}},
\end{equation}
where $N_\mathrm{He}$ and $N_\mathrm{H}$ are the number densities of helium and hydrogen, respectively. The abundances of the other elements $\epsilon_\mathrm{X}$ are defined as:

\begin{equation}
    \epsilon_\mathrm{X} = \log\frac{N_\mathrm{X}}{N_\mathrm{H}} + 12,
\end{equation}
where $N_\mathrm{X}$ is the number density of the given element. For the purposes of reducing the dimensionality of the parameter space, we use one parameter for the abundances of elements heavier than helium, setting the abundances of carbon, nitrogen, oxygen, and silicon to the same value.  This inherently assumes that changing the abundance of one element does not significantly affect the line profiles of another element, which is an acceptable assumption in most cases \citep[however, see e.g., ][for exceptions]{puls2000}. Using non-rotating evolutionary tracks from \citet{Brott2011b} as a guide, we make additional cuts to the parameter space to avoid regions where no stars are expected to be found: diagonal cuts are made in the $\log g$ versus $T_\mathrm{eff}$ space from ($T_\mathrm{eff} = $ 15~kK, $\log g = $ 2.0) to ($T_\mathrm{eff} = $ 60~kK, $\log g = $ 4.0), and in the $T_\mathrm{eff}$ versus $R$ space from ($T_\mathrm{eff} = $ 35~kK, $R = $ 30~\rsol) to ($T_\mathrm{eff} = $ 60~kK, $R = $ 5~\rsol).  With these limits in mind, samples are randomly drawn from a uniform distribution across our 5-dimensional parameter space. Our final sample consists of 50147 \fw\ models, which is illustrated in Fig. \ref{fig: param_space}.

Individual models are calculated using \fw\ v10.6.5.  Aside from the varied parameters outlined above, we set metallicity to solar and the microturbulent velocity to $v_\mathrm{micro} =$~10~\kms. We assume an unclumped wind using the \citet{Vink2001} mass loss prescription, a beta parameter of $\beta = $ 0.8 and a terminal wind speed $v_\mathrm{inf}$ derived based on the mass (which is determined based on the radius and surface gravity), radius and metallicity. All other parameters are left to their default values.  While each \fw\ line profile has exactly 161 wavelength bins, the exact wavelength values that make up these 161 points varies from model to model.  To ensure that the wavelength arrays are homogenized across the 50147 \fw\ models, we create a master wavelength array for each line by taking the median value of all of the wavelength arrays in the sample for each of the 161 points that make up the line.  Each line of each model is then interpolated onto its respective master wavelength array.  For each model, we calculate a total of 140 individual line profiles (some of which include blends), making up our master line list (see Table \ref{tab: MAEs}).

\begin{figure}
    \centering
    \includegraphics[width=0.95\columnwidth]{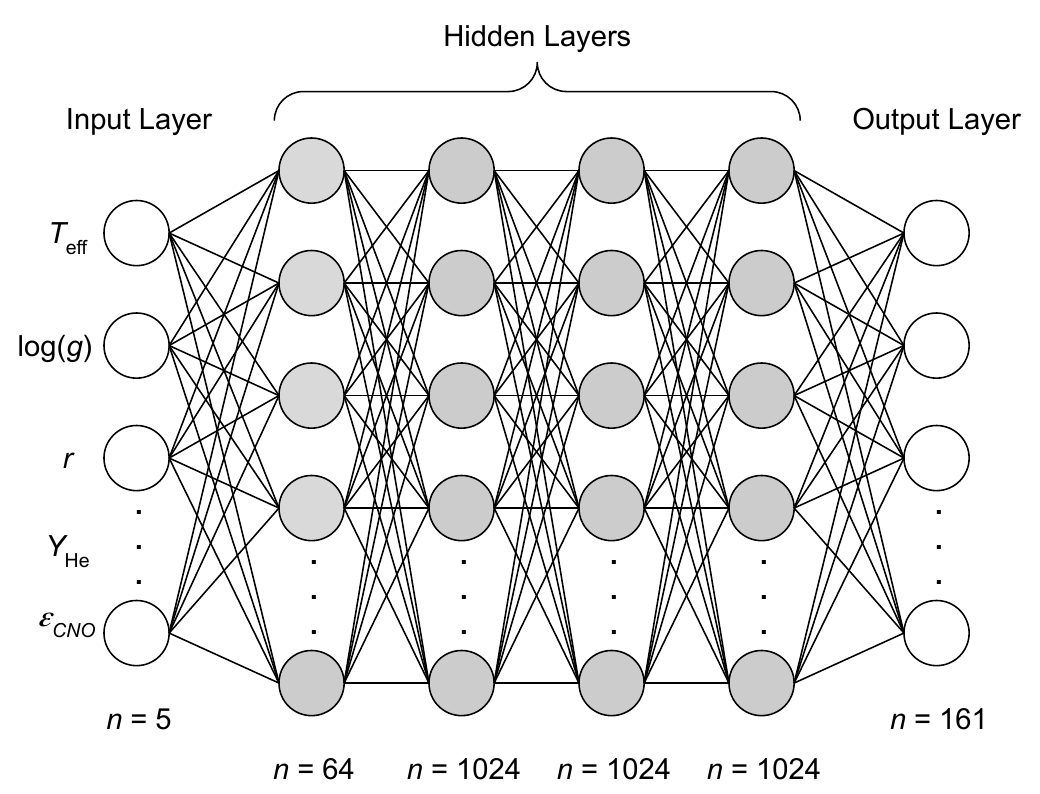}
    \caption{Schematic representation of the structure of the neural networks used for each line in our line list.}
    \label{fig: nn_architecture}
\end{figure}

\begin{figure*}
    \centering
    \includegraphics[width=0.49\textwidth]{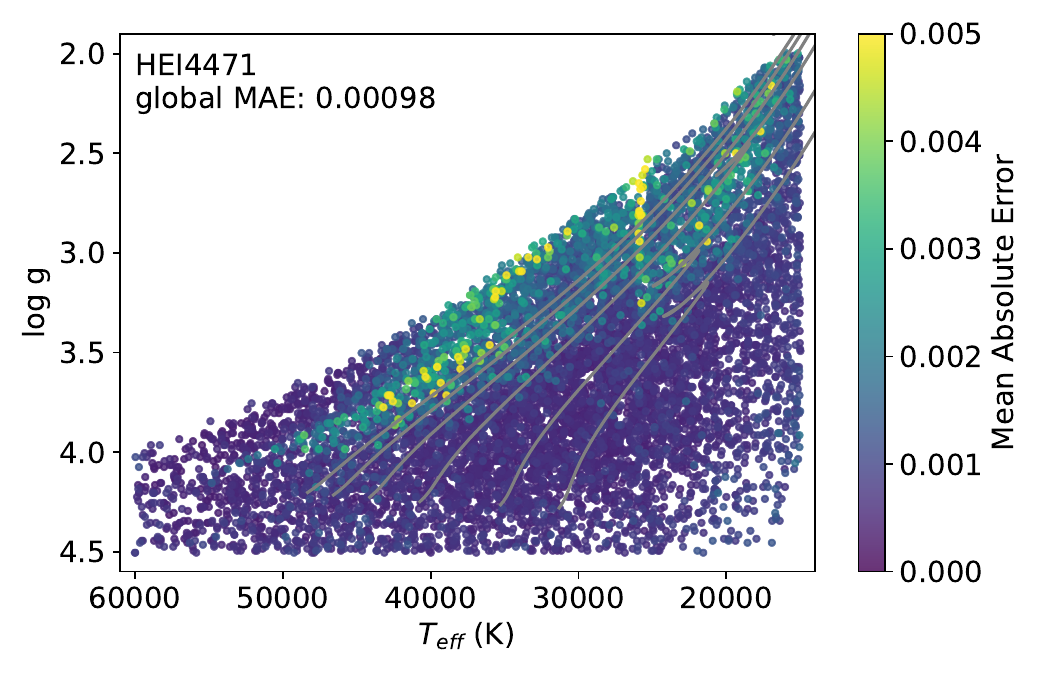}
    \includegraphics[width=0.49\textwidth]{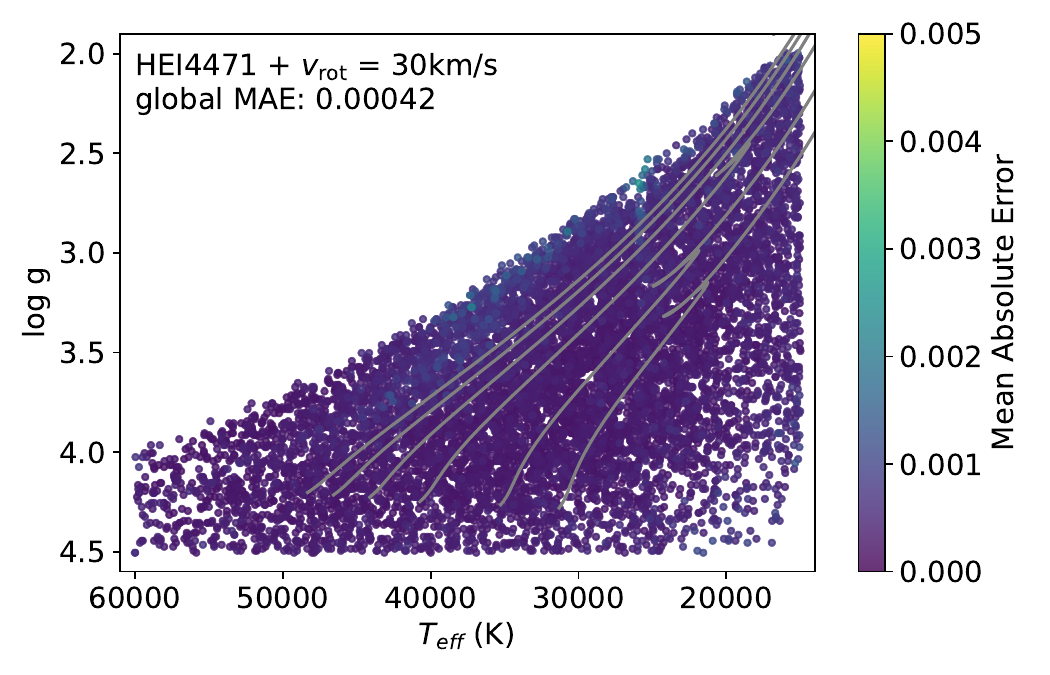}

    \includegraphics[width = \textwidth]{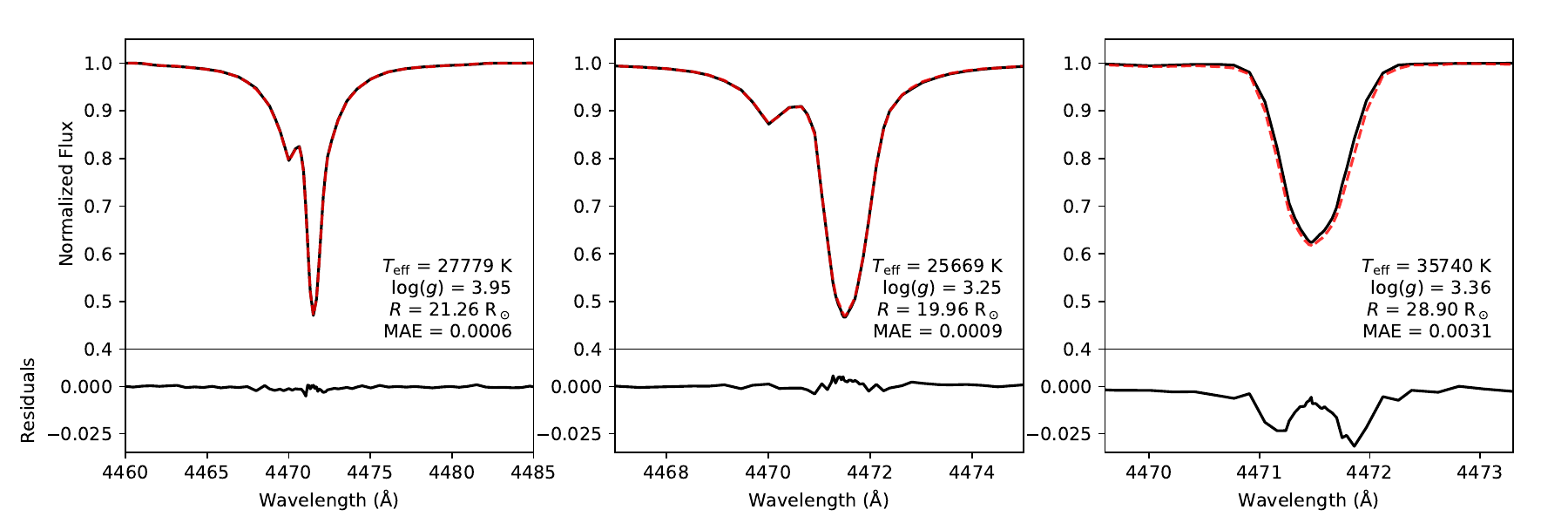}

    \caption{Top left: result of the validation of the test set for the HeI \lam4471 line.  Each point represents one model, and the symbol color shows the mean absolute error, with worse fitting models plotted on top.  Non-rotating evolutionary tracks from \citet{Brott2011b} are over-plotted in grey, with the mass of each track being the same as in Fig. \ref{fig: param_space} (60, 50, 40, 30, 20, and 15 \msun\ from left to right).  The global mean absolute error (MAE) is indicated in the top left corner. Top right: same as in the top left plot, but after convolving with a 30 \kms\ rotational kernel. Bottom: selection of model comparisons between the predictions of the neural network (red dashed lines) with the \fw\ line profiles (black solid lines) demonstrating a better-than-average fit (left), an average fit (middle), and a worse-than-average fit (right). The parameter combinations of these lines and the mean absolute error are indicated in the bottom right of each plot, and the bottom axes shows the residuals.}
    \label{fig: HeI4471_validation}
\end{figure*}

\subsection{Neural network}

For the preparation and training of the neural networks, we use \textsc{keras} \citep{Chollet2018}, which is included as part of the \textsc{tensorflow} package \citep{ abadi2016}.  We use a fully connected feed-forward neural network architecture, with separate networks for each of the line profiles in our master line list.  All networks have identical structures and hyperparameters, namely an input layer containing five nodes, four hidden layers containing 64, 1024, 1024 and 1024 nodes, respectively, and an output layer containing 161 nodes. The five nodes in the input layer correspond to the five stellar parameters that we vary, namely, \teff, \logg, $R$, \heabun, and $\epsilon_\mathrm{CNOSi}$, and the 161 nodes in the output layer correspond to the 161 wavelength points in a standard \fw\ line profile.  Each layer is fully connected with its neighboring layers, and we use the Rectified Linear Unit \citep[ReLU; ][]{agarap2018} activation function for each of the hidden layers, and a linear activation function for the output layer. A visual representation of our neural network structure can be found in Fig. \ref{fig: nn_architecture}.

We randomly divide our sample into a training set and a test set (40147 and 10000 models, respectively). We use the mean squared error (MSE) as our loss function, and we use RMSprop as our optimization algorithm \citep{hinton2012}.  For training purposes, we normalize our input data by subtracting off the mean of each parameter in the training set and dividing by the standard deviation, such that our samples are centered around 0 in each dimension.  Finally, we train each network for 20000 iterations on the training set and later validate the results of the training with the test set.  This setup corresponds to $\sim$1 hour of training time per line on an NVIDIA L40S GPU.


\section{Neural network training results} \label{sec:results}

We validate the results of the training using the test set by propagating the parameters of the test set through the trained neural network and comparing the predicted line profiles with those calculated directly with \fw. In general, we find that the MSE of the test set is quite similar to the MSE of the training set, implying that the training process is not suffering from overfitting.  While the MSE was used as our loss function during the training process, for the remainder of this work we use the mean absolute error (MAE), which corresponds to the average of the absolute value of the residuals, to discuss the results of the training as it offers a more clear and relatable comparison between the network predictions and the `ground truth' \fw\ models.

Each model in the test set is assigned an individual MAE value per line, and for each line in our master line list we obtain a global MAE, which corresponds to the average MAE for that line across all models in the test set. As an example, Fig. \ref{fig: HeI4471_validation} shows the results for the HeI \lam4471 line. In this case, the global MAE that we are able to reach is 0.00098, or $\sim$ 0.1\% agreement.  From the top left panel in Fig. \ref{fig: HeI4471_validation}, there is structure in the MAE values across the parameter space, with two notable features. The first and most obvious is at lower surface gravities around the bi-stability jump, which manifests as a vertical line of higher MAE values at around \teff~$\sim$~25kK. This arises from the choice of using the \citet{Vink2001} mass loss prescription, which contains a sharp transition where Fe ionization states induce a sharp increase in wind velocity. The second feature is less prominent, but predominantly affects models approaching the Eddington limit, manifesting as higher MAE values along the upper regions of the plot.  Comparing individual models across the parameter space also provides useful insights. The bottom row of panels in Fig. \ref{fig: HeI4471_validation} shows how the neural network predictions (red) compare to the \fw\ profiles (black) for various representative MAE values. The largest residuals are found in the line centers, which is not entirely unexpected, however, there also appears to be a noticeable `noise' throughout the entirety of the line profiles predicted by the neural network. Since rotation is common among massive stars, when a modest rotation rate of 30~\kms\ is applied to both the predictions and the models, this is enough to effectively smooth out this noise. This results in a global MAE of 0.00042, less than half of the unbroadened MAE (see the top right panel of Fig. \ref{fig: HeI4471_validation}). That being said, for stars whose total broadening (including from rotation, macroturbulence and instrumental broadening) is more than 30~\kms, the broadened global MAEs are likely more representative of the limitations of the networks.

\begin{figure}
    \centering
    \includegraphics[width=0.95\columnwidth]{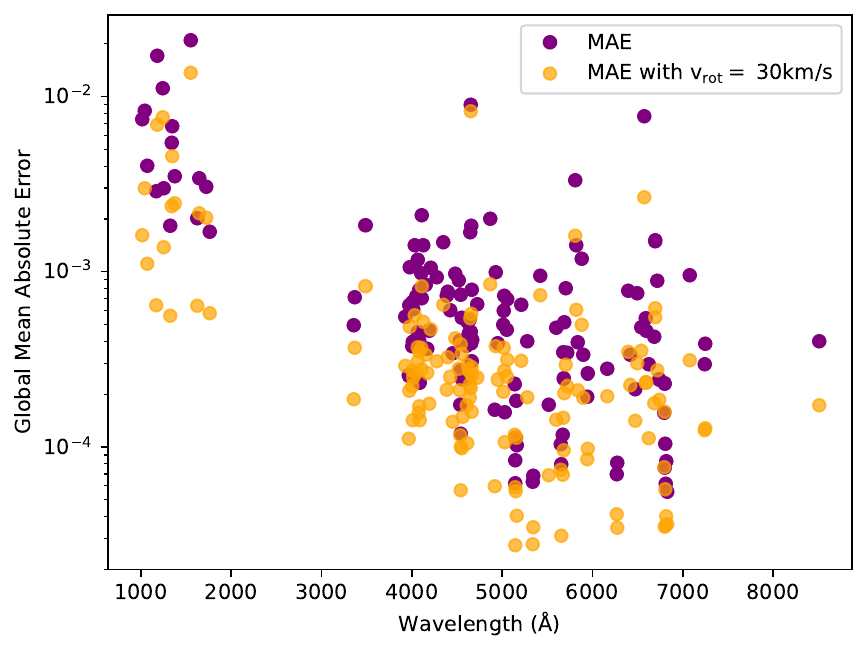}
    \caption{Distribution of global MAEs for each line in the master line list as a function of the central wavelength.  Global MAE is plotted in purple, while the global MAE after applying a 30~\kms rotation rate is plotted in orange.}
    \label{fig: global_MAEs}
\end{figure}

The global MAEs (both the unbroadened and broadened) vary by about two orders of magnitude across the 140 lines in our master line list, as demonstrated in Fig. \ref{fig: global_MAEs}.  In general, photospheric lines are reproduced better than wind lines, and among wind lines, the stronger the effect of the wind, the worse the network is able to reproduce the line. This can be seen in Fig. \ref{fig: global_MAEs}, where the UV wind lines have systematically higher MAEs than the optical lines. Similarly, the Balmer lines and \heb \lam4686 have some of the highest MAEs among the optical lines. A complete list of the unbroadened and broadened global MAE values can be found in Table \ref{tab: MAEs}.

While most of the photospheric lines show similar features in the distribution of MAEs across the parameter space, as seen in the top left panel of Fig. \ref{fig: HeI4471_validation}, the MAE distribution in the wind lines differs somewhat.  For reference, Figs. \ref{fig: HALPHA_validation} and \ref{fig: CIV1550_validation} show the same plots as Fig. \ref{fig: HeI4471_validation}, but for \halpha\ and \cd \lam1550, respectively. We see that \halpha\ shows noticeably higher MAEs close to the Eddington limit, however, the bi-stability jump feature is quite different from the one seen in the photospheric lines.  In this case, temperatures below the bi-stability jump and surface gravities below $\log g < 3.5$ show consistently higher MAEs.  In the case of \cd \lam1550 (the line with the worst MAE in our master line list), temperatures above the bi-stability jump appear to show consistently higher MAEs.  That said, in both cases, despite the high MAE values, the networks can still reproduce the complex features quite well, with the main source of error originating from the strength of the wind feature more so than the shape.   

\section{The \specfann\ python package} \label{sec:specfan}
Now that we have created a high-fidelity \fw\ emulator, we can begin to explore how this can be applied to fitting observed data. Because we have trained separate networks for each individual line (henceforth, we refer to the collection of networks as `the bundle'), we can essentially use these networks as drop-in replacements for \fw, allowing us to use many of the post-processing and fitting techniques that have already been developed and tested without needing to modify them in any significant way.  Additionally, due to the significantly reduced run times when compared to \fw\ (see Sec. \ref{sec:discussion}), we can begin to explore alternative fitting techniques that were not feasible in the past.  To facilitate the interaction with the bundle of networks, we have developed the \linebreak \specfann\footnote{\url{https://github.com/MichaelAbdul-Masih/SpecFANN}}\ (Spectral Fitting via Artificial Neural Networks) Python package.  In addition to providing convenience functions for converting input stellar parameters to line profiles, \specfann\ also provides a number of post-processing functions, cost functions, and fitting methods, allowing a user to easily fit observed spectra with the customized set-up and line lists that fit their specific use case.   

The parameter range over which \specfann\ can be reliably used is defined by the parameter space covered by the training set.  It is important to note that while forward models can be generated for parameter combinations that fall outside the parameter space covered by the training set, these should be used with extreme caution as their fidelity cannot be guaranteed.  Similar to \fw, the neural networks return unbroadened line profiles at laboratory wavelengths, so we provide broadening functions to account for rotation, macroturbulence, and instrumental resolution, and a Doppler shift function to account for non-zero radial velocities. \specfann\ allows the user to generate line profiles for either an individual star or for a sample of stars simultaneously. Given the vectorized nature of neural networks, the computation time scales sublinearly with the size of the input, meaning that the efficiency increases when more stars are modeled simultaneously.  All the broadening functions provided are also vectorized to minimize runtime as much as possible.

At the time of writing, four fitting methods are included within \specfann: the Nelder-Mead algorithm, a genetic algorithm (GA), Markov Chain Monte Carlo (MCMC) and Nested Sampling (NS).  For the Nelder-Mead algorithm, we use the \texttt{optimize.minimize} function included in the \scipy\ package \citep{Virtanen2020}.  For the GA, we use the \pyGA\ package \citep{Abdul-Masih2021}, for MCMC we use the \emcee\ package \citep{Foreman-Mackey2013}, and for NS we use the \ultranest\ package \citep{buchner2016, buchner2019, buchner2021}.  Additionally, \specfann\ provides different loss functions that can be used by each of these fitting methods, including chi-square, log likelihood, and log probability. \specfann\ allows the user to fix and free any of the parameters included in the parameter set and to customize the bounds of each of the free parameters. The user can also select the specific lines they wish to include in their line list, and the fitting bounds of each line can also be customized.  The line list and parameter set are standardized across each of the different fitting methods to allow for a seamless transition between the different methods.  Additionally, \specfann\ provides built-in plotting functions to display diagnostic plots and line profile fits for each of the included fitting methods.

\begin{table}
\caption{Derived stellar parameters for 10 Lac.}
\centering 
\renewcommand{\arraystretch}{1.5}
\resizebox{\columnwidth}{!}{
\begin{tabular}{cccc}
\hline\hline
 & GA & MCMC & \citet{Holgado2025} \\
\hline
\teff\ (kK)       & $35.8^{+0.3}_{-0.4}$   & $36.06^{+0.04}_{-0.05}$ & 35.4 $\pm$ 0.7\\ 
\logg             & $3.97^{+0.01}_{-0.04}$ & $3.994^{+0.009}_{-0.012}$ & 3.93 $\pm$ 0.09 \\
\vsini\ (\kms)    & $33^{+4}_{-5}$        & 32.5 $\pm$ 0.4 & 13 \\
$\gamma$\ (\kms)  & $-9.6^{+1.7}_{-1.5}$         & $-10.85^{+0.14}_{-0.13}$ & -10.5 $\pm$ 0.8 \\
\heabun           & $0.11\pm0.01$        & $0.1063 \pm 0.0013$ & 0.10 $\pm$ 0.02 \\
\cabun            & $8.34^{+0.13}_{-0.09}$           & $8.396^{+0.012}_{-0.011}$ & - \\
\nabun            & $7.85\pm0.12$    & $7.921^{+0.016}_{-0.014}$ & - \\
\oabun            & $8.6^{+0.3}_{-0.9}$                      & $8.56^{+0.03}_{-0.02} $ & - \\
$\log f$          & -                      & $-3.085 \pm 0.003$ & - \\
 \hline
\end{tabular}
}
\tablefoot{1$\sigma$ confidence intervals for fits from the GA, MCMC and from \citet[]{Holgado2025} are indicated.  Parameters not included in the respective fits are indicated by `-'.}
\label{tab: 10lac}
\end{table}

\section{Example application of \specfann} \label{sec:application}

\subsection{Fitting 10 Lac} \label{sec:10lac}
To validate that the neural networks and the \specfann\ implementation can obtain accurate stellar parameters, we apply \specfann\ to the well-studied standard O-type star 10 Lac \citep[HD\,214680; e.g., ][]{Royer2024, GiudiciMichilini2025, Holgado2025}.  We use a high-resolution spectrum from the HERMES spectrograph mounted on the Mercator Telescope \citep[$\mathcal{R}~=~d\lambda/\lambda$~=~85000, $\lambda \approx3800 - 9000$~\AA; ][]{Raskin2011} retrieved from the IACOB database \citep{Simon-Diaz2014}, and we calculate the error per point based on the standard deviation of a continuum region of the observed spectrum.  We select a line list that includes 16 lines (some of which contain blends), including two Balmer lines, four \hea\ lines, three \heb\ lines, two \cc\ lines, three \nc\ lines, one \nd\ line, and one \oc\ line. The fitting ranges of each line are left to their default values, which correspond to the middle 50\% of the total extent of the line in wavelength space as modeled by \fw. We fix the stellar radius to $R = $ 7.3~\rsol\ following the calibrations in \citet{Holgado2025}, and we fix the silicon abundance to \siabun\ $ = $ 7.0, the macroturbulent velocity to \vmac~$= 10$~\kms\ and the instrumental resolution to $\mathcal{R}$~=~85000.  All other parameters (\teff, \logg, \heabun, \cabun, \nabun, \oabun, \vsini, and the radial velocity $\gamma$) are left free. We perform three separate fits, the first with the GA, the second with MCMC, and the third with NS.

\begin{figure*}
    \centering
    \includegraphics[width=\textwidth]{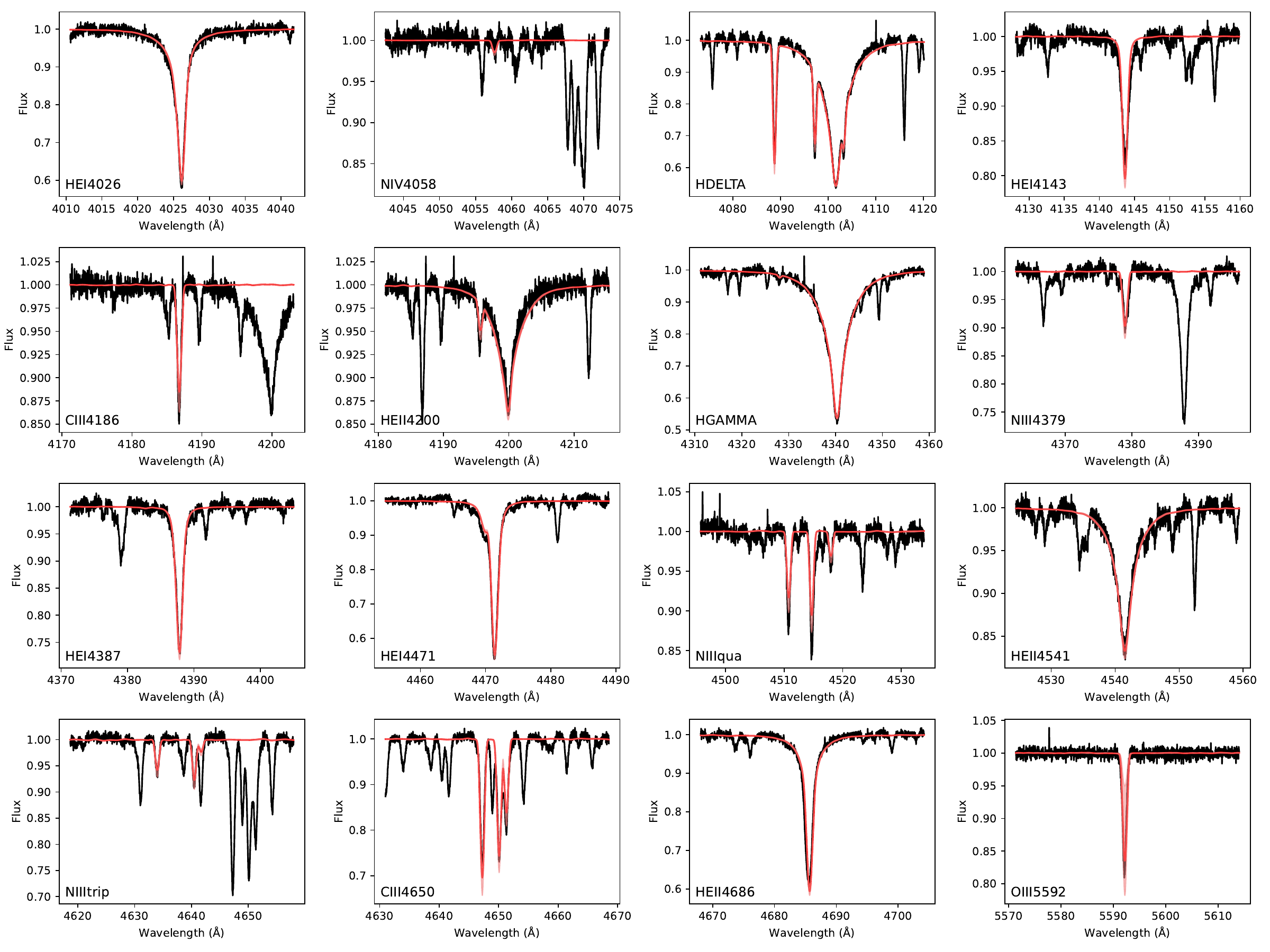}\\
    \vspace{5mm}
    \includegraphics[width = \textwidth]{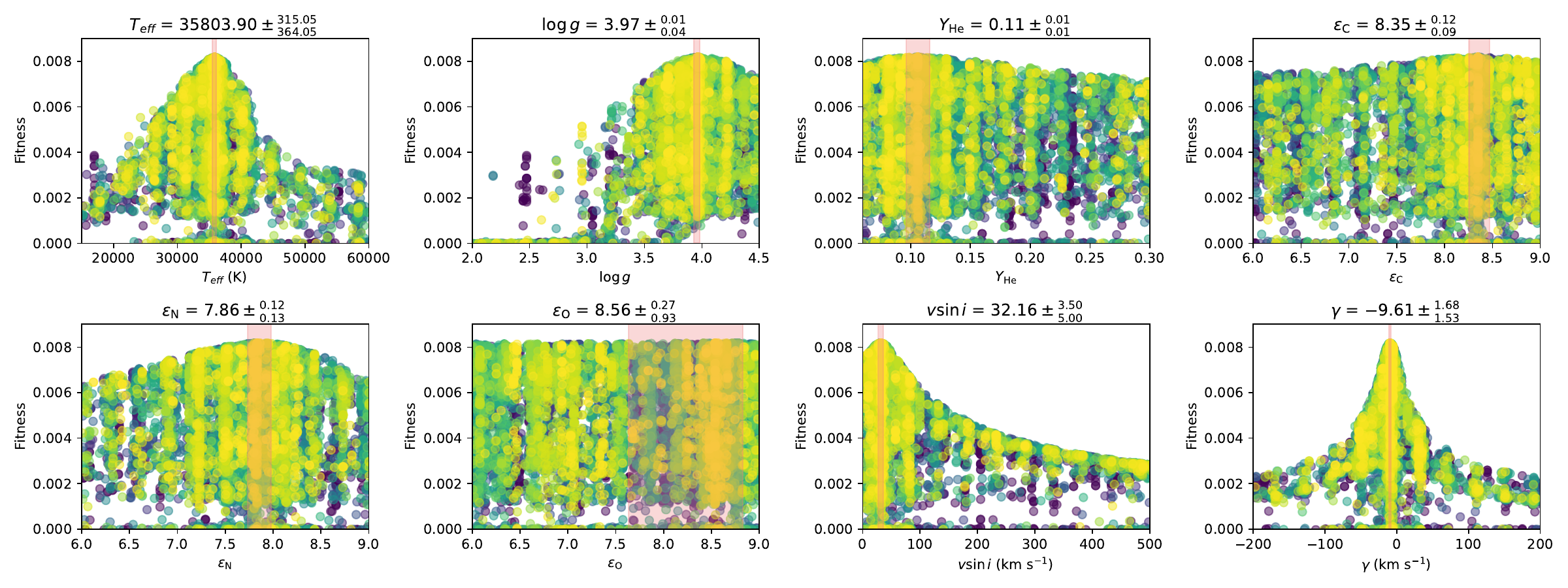}

    \caption{Results of the GA fit for 10 Lac.  The top panels show each of the 16 fitted lines with the observed spectrum plotted in black and the synthetic spectrum of the best-fitting model plotted in red.  The shaded red regions represent all models that fall within the 1\s\ confidence interval.  The bottom panels show the fitness as a function of each of the free parameters in our parameter set.  The color of each point represents which generation the model belongs to with darker blue colors representing older generations and lighter yellow colors representing more recent generations. The 1\s\ confidence interval is represented with the red shaded region.  The best-fit solution and 1\s\ confidence interval for the respective parameter is indicated above each panel.}
    \label{fig: GA_fit}
\end{figure*}

\begin{figure*}
    \centering
    \includegraphics[width=\textwidth]{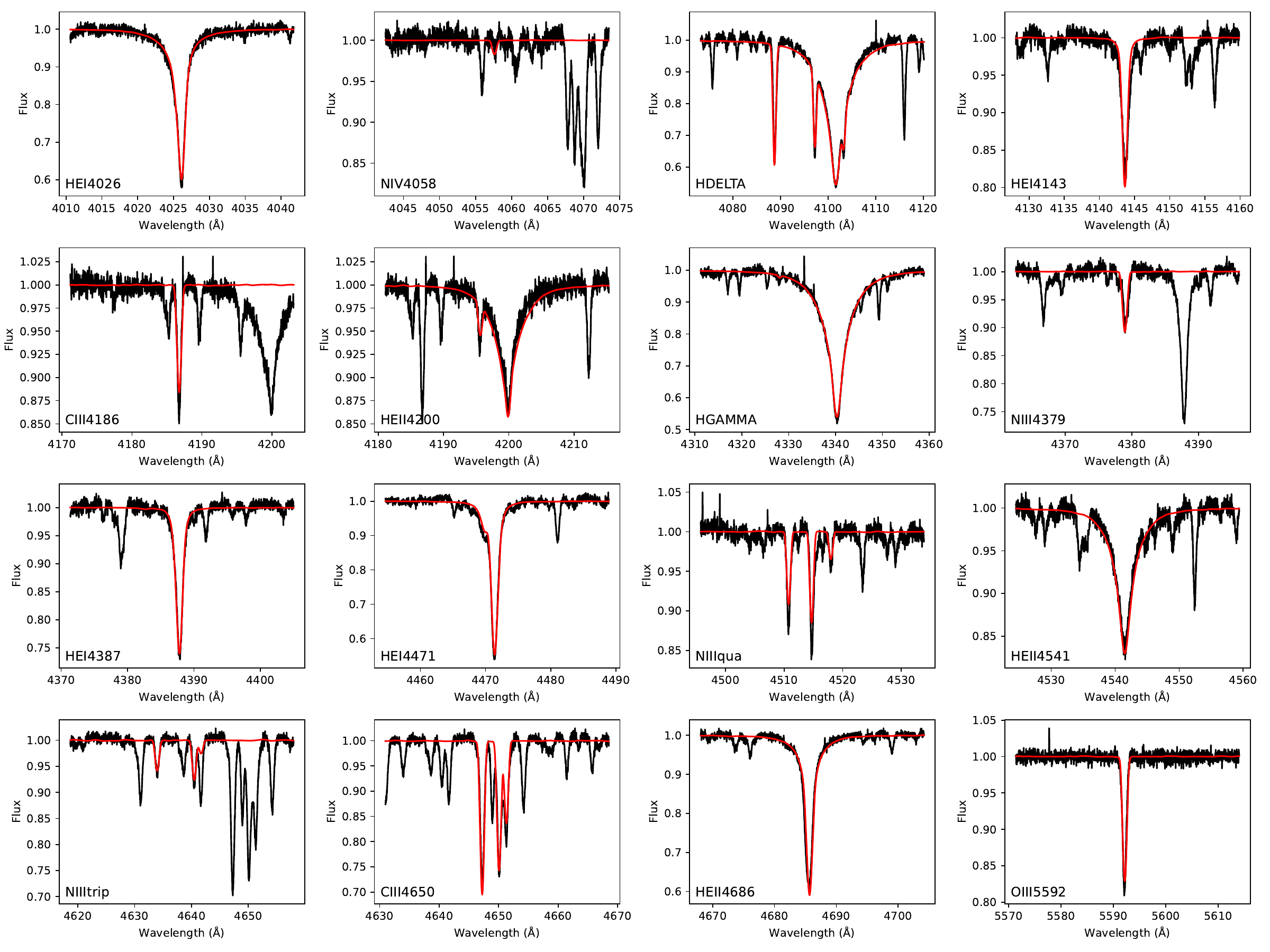}

    \caption{Results of the MCMC fit for 10 Lac.  This figure is the same as the top panels in Fig. \ref{fig: GA_fit}, but for the results of the MCMC instead.}
    \label{fig: MCMC_fit}
\end{figure*}

\subsubsection{Fitting with the Genetic Algorithm}
Using the line list and parameter setup outlined above, we run the GA for 300 generations with a population size of 50.  The results of the GA analysis are shown in Fig. \ref{fig: GA_fit}.  The bottom panels of Fig. \ref{fig: GA_fit} show that each parameter was well explored, and that the shape of the manifold is well defined and clear.  This, in combination with the high quality of the line fits in the upper panels, indicates that the global best-fit solution has indeed been found and is reliable.  As shown in Table \ref{tab: 10lac}, the best-fit solution that we find here matches very well with the values reported in \citet{Holgado2025}, who fit using a grid of \fw\ models.  Specifically, our best-fit values for \teff, \logg, $\gamma$, and \heabun\ all fall within the error ranges reported in \citet{Holgado2025}.  The only parameter that appears in both of our fits that is inconsistent is the \vsini, however this is expected as our \vsini\ determination methods are different.  \citet{Holgado2025} determines the \vsini\ using the Fourier method \citep{Simon-Diaz2007}, which assumes that the first zero in the Fourier transform of a spectral line corresponds to the \vsini. This is then fixed for the rest of their analysis.  In our case, we include \vsini\ as a free parameter in our fit.

\subsubsection{Sampling the posteriors with Markov Chain Monte Carlo}
\begin{figure*}
    \centering
    \includegraphics[width = \textwidth]{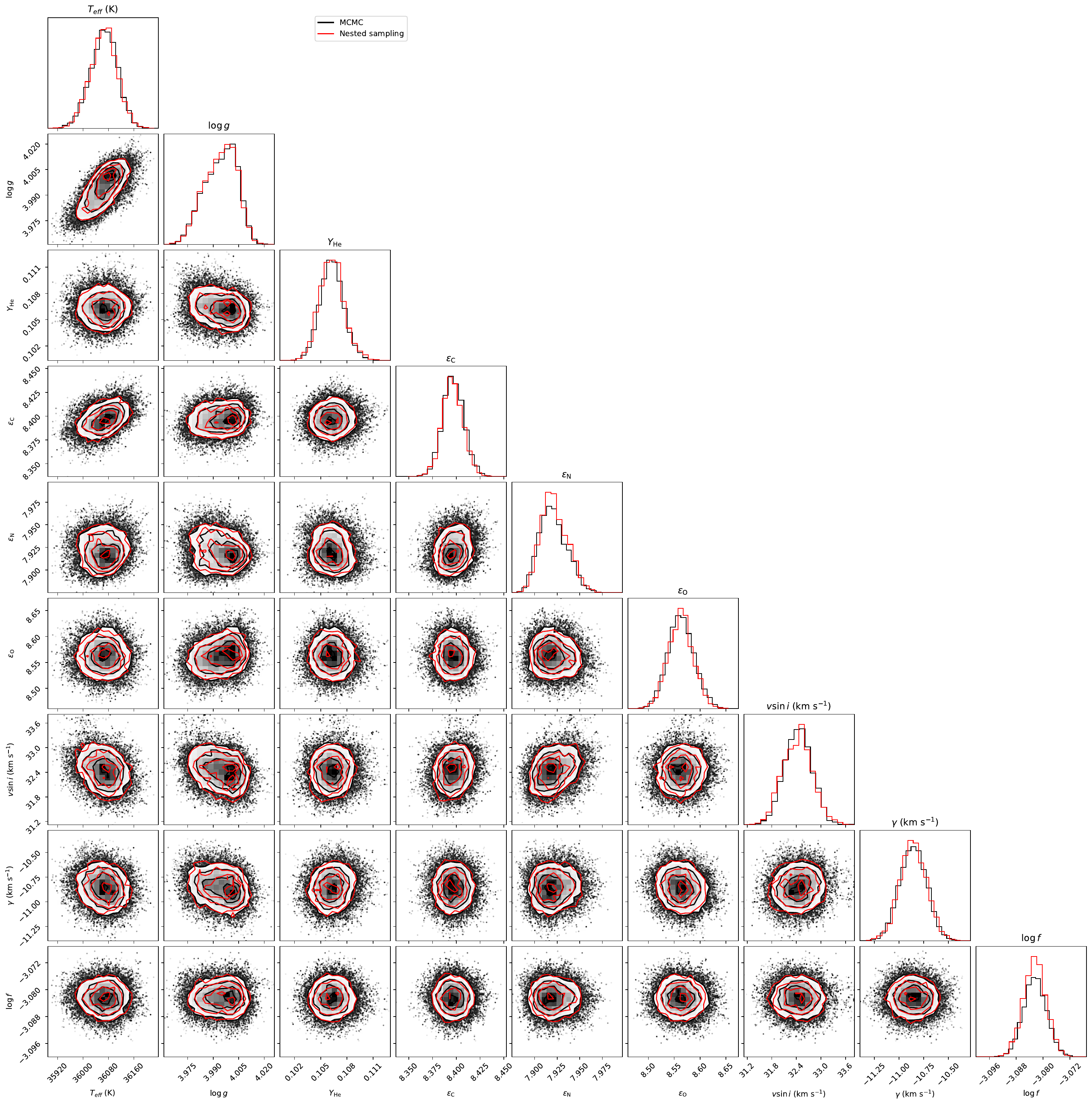}

    \caption{Corner plot showing the posterior distribution and contours of the MCMC fit in black with the Nested Sampling contours overplotted in red for 10 Lac.  The top panel in each column shows the distribution for the parameter indicated above the plot, and the off-diagonal panels show the parameter correlations.  }
    \label{fig: MCMC_NS_corner}
\end{figure*}

For the MCMC fit, we use the same line list and parameter set as with the GA, however we also include a fuzz term ($\log f$) as a free parameter, which accounts for underestimated per-point errors in the observed data. This is due to our choice of cost function: while we use $\chi^2$ for the GA fit, in the case of MCMC we use log probability, which includes $\log f$ in its formulation. We assume uniform priors and all parameters are initialized randomly within the bounds of the parameter space.  We run the MCMC with 50 walkers for 5000 iterations.  Figures \ref{fig: MCMC_fit} and \ref{fig: MCMC_NS_corner} show the results of the fitting after determining and discarding a burn-in of 1000 iterations.  As with the GA results, the MCMC line profile fits are of high quality, indicating that we have successfully reached the global minimum.  The posterior distributions are Gaussian with minimal correlations outside the known correlations between \teff\ and \logg.  While there are slight differences in some of the parameters between the GA and MCMC, they are consistent with one another \citep[and with][ see Table~\ref{tab: 10lac}]{Holgado2025}.  While the GA was not able to tightly constrain the oxygen abundance, our MCMC framework was even though only a single \oc\ line was included in the analysis.

\subsubsection{Sampling the posteriors with Nested Sampling}
Our NS setup mirrors the MCMC setup described above.  We use the same line list, parameter set, cost function, and priors as used in the MCMC fit.  For the hyperparameters, we use a minimum of 500 live points drawn with the \texttt{RobustEllipsoidRegion} regions and default exploration. Compared to MCMC, NS explores the full prior volume (rather than a single local posterior mode) and additionally yields an estimate of the Bayesian evidence for model comparison. Figure \ref{fig: MCMC_NS_corner} shows the contours of the posteriors of the NS fit in red, and since it is so similar to the MCMC posteriors, we do not show fit here as it is qualitatively and quantitatively identical to the MCMC fit shown in Fig. \ref{fig: MCMC_fit}. As with MCMC, NS was also able to constrain the oxygen abundance.

\begin{figure*}
    \centering
    \includegraphics[width=0.52\textwidth]{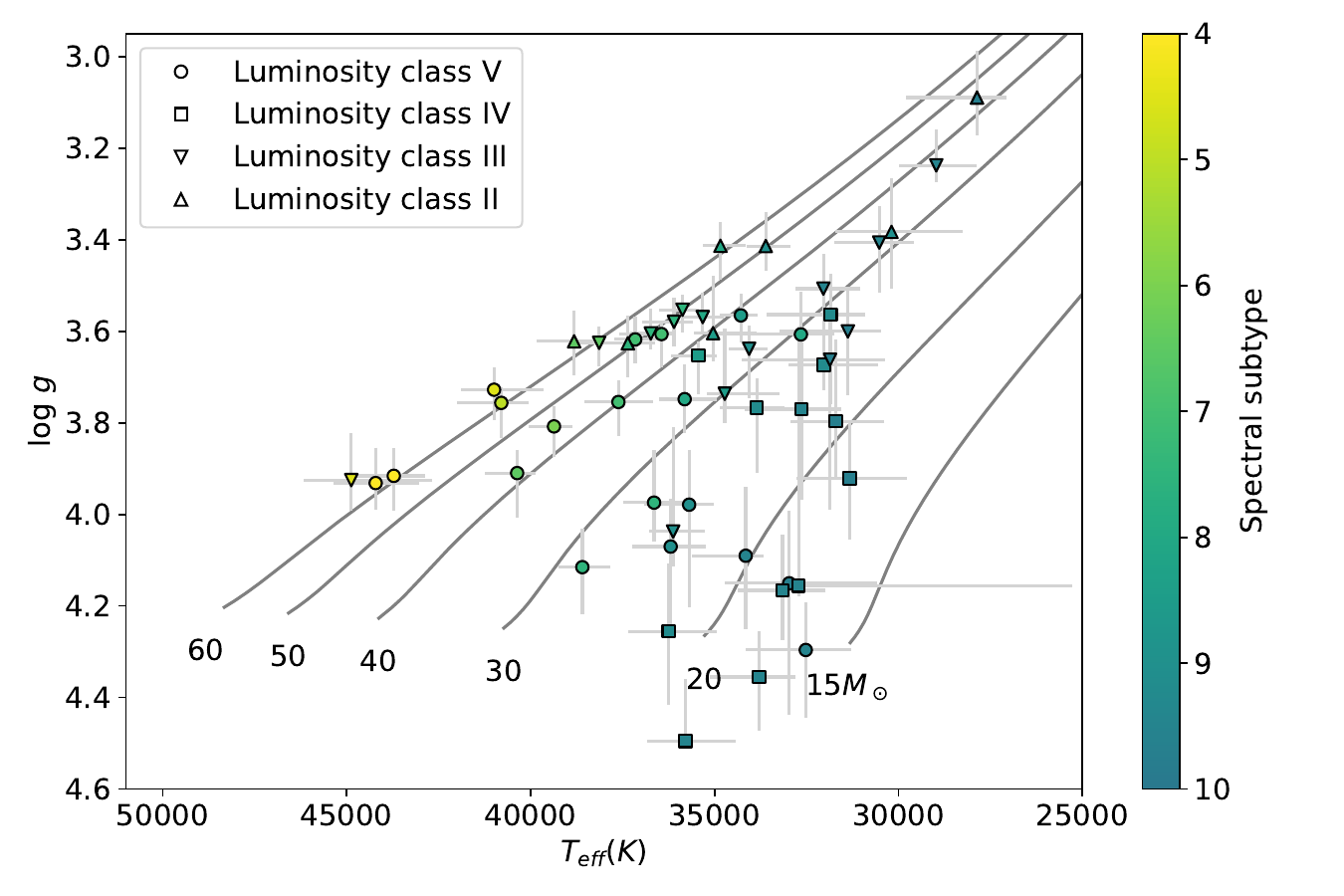}
    \includegraphics[width=0.46\textwidth]{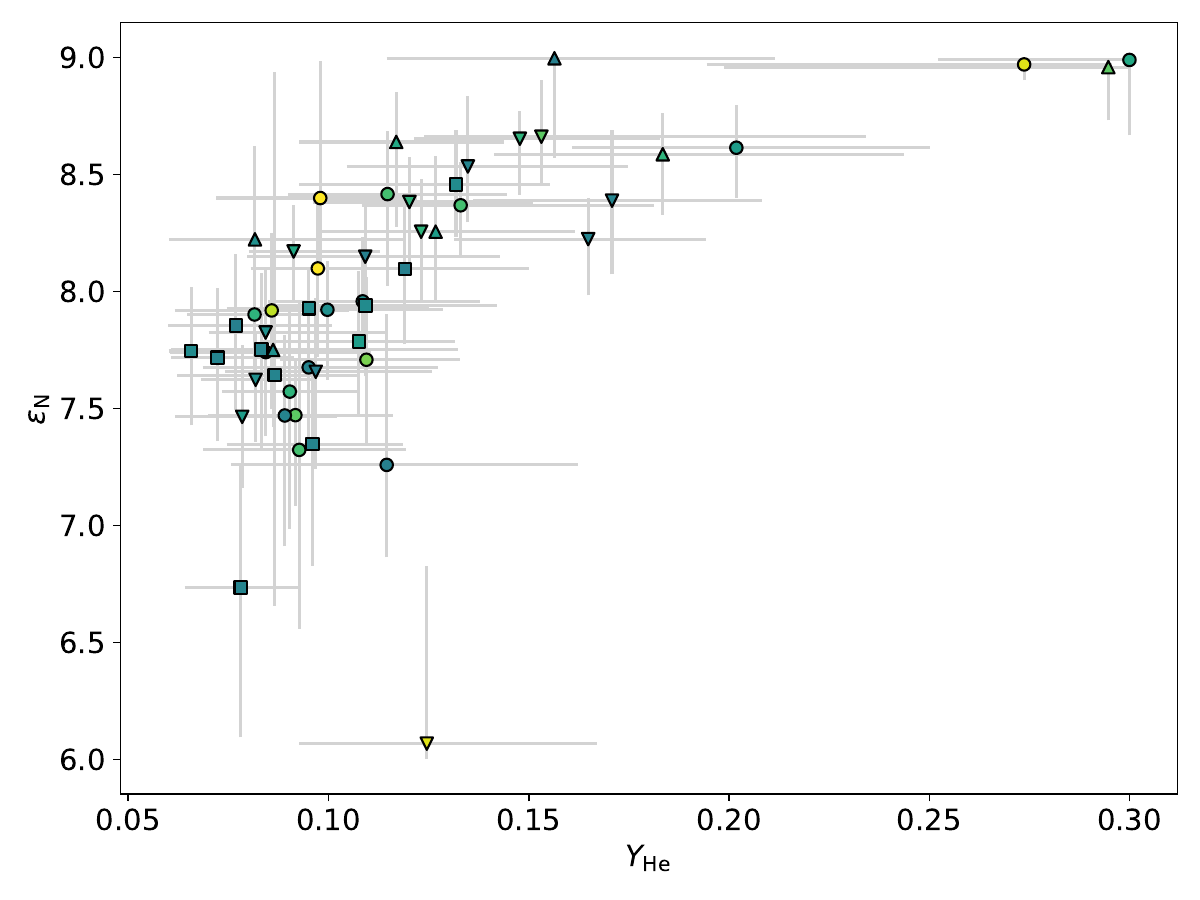}

    \caption{Results of the GA fit for the single O-type stars in the Melchiors database in a Kiel diagram (left) with \citet{Brott2011a} tracks plotted in dark grey, and in the nitrogen abundance versus helium abundance plane (right). Error bars are indicated in light grey, spectral subtypes are indicated by the color of each point, and the luminosity classes are indicated by the symbol type as indicated in the legend of the left panel.}
    \label{fig: melchiors}
\end{figure*}

\subsection{Single O-type stars in the MELCHIORS Database}
With \specfann\ validated using the example of 10 Lac, we now apply \specfann\ to a collection of stars across a broad range of the parameter space that the bundle is sensitive to.  To do this, we use the MELCHIORS database \citep{Royer2024}, which provides more than 3000 reduced, automatically-normalized, high-resolution HERMES spectra of spectral standard stars. We select all of the O-type stars marked as spectroscopically single in the database, removing any object with an uncertain spectral type (i.e., ranged subtype or missing luminosity class).  To remain within the parameter space that the bundle was trained on, we remove supergiants from the sample. This results in a total of 52 stars covering spectral types from O9.7 to O4 with luminosity classes of II-V \citep[for a breakdown of the O-type sample, see Fig. 12 of ][]{Royer2024}.  The radii are set based on the calibrations from \citet{Holgado2025}, and we fit the sample with the GA using the same set up and line list as described above.

In general, the fits are reasonable with fit qualities similar to that achieved for 10 Lac using the GA.  In many cases, however, the quality of the data itself is worse (lower S/N, poorer normalization, etc.), which is reflected in larger error bars.  Based on the reported spectral types and luminosity classes, the values of temperature and surface gravity that we obtain are in line with expectations. This can be seen in the left panel of Fig. \ref{fig: melchiors}, which shows the 52 stars plotted in a Kiel diagram. Here, we demonstrate that \specfann\ can obtain reliable fits across the entire O-star regime, and that there is no preference towards specific combinations of parameters during the fitting process.  The right panel of Fig. \ref{fig: melchiors} shows the nitrogen abundance as a function of the helium abundance, demonstrating the expected correlation resulting from CNO cycle processing.

\section{Discussion} \label{sec:discussion}

\subsection{Neural network performance} \label{subsec:net_performance}
Overall the agreement between the trained neural networks and the reference \fw\ models is excellent.  In most cases, the residuals that arise from differences between the networks and \fw\ would be significantly below the noise level in a given spectrum, meaning that the errors in a fit would be dominated by the signal-to-noise of the observed spectrum as opposed to the quality of the trained neural network.  In the worst cases (corresponding to UV wind lines), the MAEs reach the $\sim$1\% level, however, our choice to train individual neural networks for each line plays in our favor.  Not only can we tailor our chosen line list to the specific observed spectrum when fitting, but we can also focus on improving the performance of the neural networks for these specific lines without altering the performance of other lines.  As long as the same training and test sets are used, alternative neural network architectures can be implemented for these problematic lines, and these newly trained networks can seamlessly replace the original networks when fitting. This flexibility also means that we could use less complex architectures for well-behaved lines, reducing both run time and storage space needed to store these models. 

While on the surface these neural networks may appear to be black boxes, it is useful to think of them in the same way that one thinks about a pre-computed grid of \fw\ models.  Both are designed to cover specific areas of the full parameter space, and both are limited by the assumptions and the specific (uncertain) prescriptions for physical processes at the time that they were computed.  In the same way that one should not extrapolate outside of the range of a grid, one should avoid extrapolating outside of the parameter space that the bundle was trained on.  Similarly, neither interpolating within a grid nor calculating a model with a neural network will ever be as accurate as computing a model with \fw\ directly. Additionally, as new physics is implemented in \fw, bugs are corrected, or atomic data is updated, the models that make up both grids and the training and test sets for the bundle will need to be rerun.  That said, with these caveats in mind, both methods offer the same benefit over calculating \fw\ models on the fly: after an initial computational investment, both allow for a significant reduction in computation time down the line.

With all of this in mind, once properly trained, neural networks offer an attractive alternative to traditional grids. With the same number of models, a trained network can cover a much larger parameter space, can reproduce \fw\ models with smaller errors, and is significantly lighter in terms of storage space.  If we consider the parameter space covered by the networks presented here, covering the same 5 dimensions using a traditional regular grid with only 10 points in each dimension would result in a grid of 100\,000 models, double what was used for the bundle. The sampling along each dimension would be entirely insufficient to allow for accurate interpolation. At the same time, a grid of 50\,000 \fw\ models would require $\sim$ 750~GB of storage space while the entire bundle of neural networks for the same set of lines requires $\sim$ 2~GB of space with the current network architecture.

\subsection{SpecFANN performance}

As demonstrated in Sect. \ref{sec:application}, \specfann\ is not only able to produce convincing fits of observed spectra, but it can also recover the stellar parameters reported in the literature to a high degree of accuracy.  This further validates the robustness of the trained neural networks presented here. The errors that we obtain with the GA are similar to those presented in the literature, indicating that inaccuracies in the networks do not significantly contribute to the overall error. That said, the neural network implementation in \specfann\ offers a significant speed-up to the fitting process when compared with traditional techniques.  A \fw\ model that includes hydrogen, helium, carbon, nitrogen, oxygen, and silicon as explicit elements takes on the order of an hour to compute, meaning that if one wanted to run the GA using the same setup as described in Sec. \ref{sec:application}, this would require $\sim$15\,000 CPU hours to complete. Using \specfann, we can run the same fit on a single core in $\sim$ 2.5 min (these computations were done on an AMD Ryzen 7 9800X3D CPU).  This represents a speed-up factor of $\sim 3.6 \times 10^5$, which also implies a reduction in the carbon footprint by the same amount during the fitting process. Currently, the bottleneck in our fitting procedure lies in the convolutions required for instrumental, rotational and macroturbulent broadening, not the propagation through the neural networks, meaning that with further optimization, we can increase the speed even further. Such drastic reductions in computation time mean that sampling methods like MCMC and NS are now viable: the MCMC run and the NS run presented in Sec. \ref{sec:application} took just over an hour each on a single core. That said, while the run times for MCMC and NS are comparable to one another in this example, altering the dimensionality and extent of the parameter space in the case of NS can lead to further reductions in computation time, while this may not necessarily be the case for MCMC.

With the large amounts of data expected in the coming years, such speed-ups will prove necessary to ensure that we are not limited by the speed of our fitting methods.  As an example, there are roughly 50\,000 massive stars with spectra available in \mbox{SDSS-V}.  With a 10 node cluster with 50 cores per node, fitting the entire sample with the GA using \fw\ would take approximately 342 years.  Using the same cluster with \specfann, we could perform these GA fits in $\sim4.2$ hours, and we could run MCMC on the full sample with 5000 iterations per run in just over 4 days. Indeed, the analysis of this sample is already underway and will be presented in a future publication.

From a programming standpoint, \specfann\ is written in a modular and flexible way, allowing new fitting techniques, loss functions, and post-processing functions to be easily added by the user. Additionally, \specfann\ is specifically designed with alternative bundles in mind, providing a straightforward way for a user to use their own custom networks while still taking advantage of the full fitting infrastructure developed here. We provide methods for the user to specify new parameters, new priors, and new lines if their networks require them without needing to alter the main \specfann\ code.  

\subsection{Comparison between fitting methods}

In Sect. \ref{sec:10lac} we used \specfann\ to fit 10 Lac with three different fitting methods. While the derived parameters from the three fits were consistent with one another within their respective uncertainties, the scales of these uncertainties differ by almost an order of magnitude (see Table \ref{tab: 10lac}). A direct comparison between these may not be warranted as the uncertainty determination methods differ, however, it certainly warrants discussion and careful consideration.  It is important to note that here we are only accounting for fitting uncertainties and that we do not include any systematics in this analysis.

In the GA fit we used $\chi^2$ as our cost function, and our uncertainty determination was done by first normalizing the $\chi^2$ values such that the best fitting model had $\chi^2_\mathrm{red}=1$.  These $\chi^2_\mathrm{red}$ were then converted to probabilities using the incomplete gamma function and uncertainties were determined by including all models whose probability were larger than a threshold value.  In this case, we choose threshold values of 0.32, roughly corresponding to a 1$\sigma$ confidence interval (however, note that without fully sampling the posterior, this is only an approximation).  

Our MCMC and NS fits both used the same cost function (log probability) and uncertainty determination method.  The log probability is calculated by adding the log priors (which are flat) and the log likelihood ($\ln\mathcal{L}$), which is defined as follows:

\begin{equation}
    \ln\mathcal{L} = -\frac{1}{2}\sum\limits_{n}\left[ \frac{\left(O_n-C_n\right)^2}{s_n^2} + \ln(2\pi s_n^2)\right],
\end{equation}

\noindent where $O_n$ and $C_n$ are the observed and calculated fluxes and $s_n^2$ is defined as:

\begin{equation}
    s_n^2 = \sigma_n^2 + f^2C_n^2,
\end{equation}

\noindent where $\sigma_n$ is the per point observational errors and $f$ is the fuzz term described above.  In the case where the fuzz term is not used, the second term in the log likelihood equation is omitted and the $s_n^2$ is set to $\sigma_n^2$, resulting in a log likelihood of:

\begin{equation}
    \ln\mathcal{L} = \frac{-\chi^2}{2}.
\end{equation}

The final uncertainties are calculated based on the posterior distribution.  Since both MCMC and NS use the same uncertainty determination method, they can be directly compared, and as seen in Fig. \ref{fig: MCMC_NS_corner}, the posterior distributions are nearly identical.

When comparing the uncertainties from the GA with those from MCMC and NS, the GA uncertainties are about a factor of 10 larger in most cases; however, it is unclear whether this results purely from the differences in the uncertainty determination methods or this reflects the effectiveness of the fitting methods themselves.  To test this, we analyze a set of 38 independent observations of 10 Lac taken with the FIES instrument \citep[$\mathcal{R}=46000$; ][]{telting2014} with both the GA and with MCMC. Since MCMC and NS resulted in consistent posteriors for this object (see Fig. \ref{fig: MCMC_NS_corner}, we only consider one of the two methods here.  The observations were acquired at different times, and were each reduced and normalized independently of one another.  Using the same setup as described in Sect. \ref{sec:application}, we fit each observation with both methods. 

\begin{figure}
    \centering
    \includegraphics[width=0.49\textwidth]{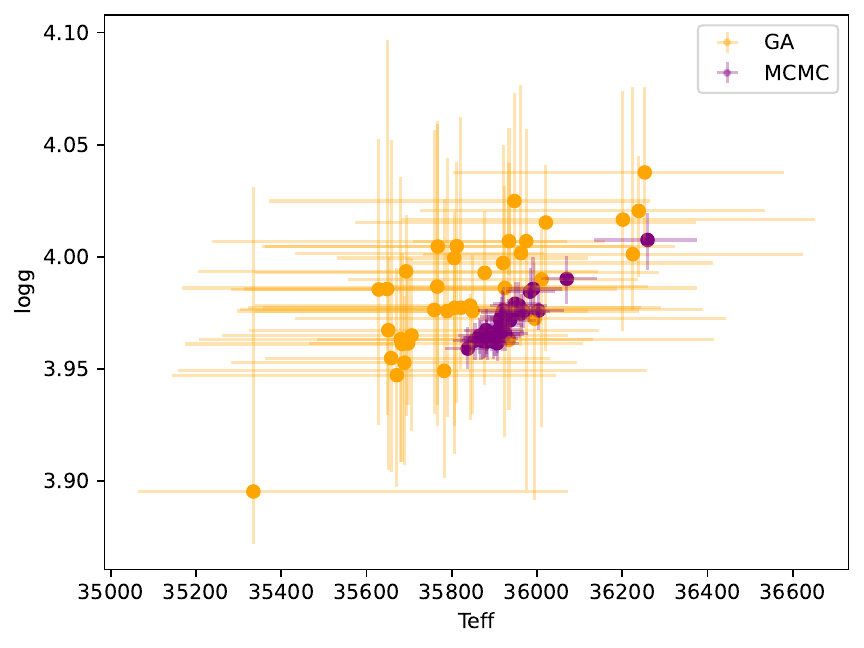}

    \caption{Comparison between the results of the GA fits (orange) and MCMC fits (purple) for 38 independent observations of 10 Lac. }
    \label{fig: ga_mcmc_comparison}
\end{figure}

The results of the fits can be seen in Fig. \ref{fig: ga_mcmc_comparison}, with the GA fits plotted in orange and the MCMC fits plotted in purple.  From this figure, it is immediately clear that the best fit values from MCMC show a significantly lower variance (i.e., tighter clustering) than the GA fits. Since each observation was reduced and normalized separately, some scatter in the final values is expected, however, this inherent scatter should contribute equally to both methods.  The fact that MCMC is able to repeatedly reproduce consistent parameter values with minimal scatter indicates that its inferred uncertainties are well aligned with the empirical variance, whereas the GA results show substantially larger dispersion and less consistency between fits, which is also reflected in its larger uncertainties.

It is likely that this difference is caused by a combination of both the fitting method and the uncertainty determination method. In the case of MCMC, the uncertainties are comparable in scale to the variance across the independent observations, indicating that the posterior distributions provide a realistic description of the parameter uncertainties.  On the other hand, in the case of the GA, the uncertainties appear to be larger than the empirical scatter, suggesting that they are systematically overestimated.  This is likely caused by the normalization of the $\chi^2$ values (forcing $\chi^2_\mathrm{red} = 1$), which effectively rescales the posterior surface and can artificially broaden the inferred confidence intervals. Additionally, the tight clustering and small uncertainties in the MCMC results indicate that the posterior distribution is quite narrow in the region of the highest probability.  Given the scatter in the GA results, it is plausible that the GA may be unable to resolve the curvature of multi-dimensional posterior distribution near the optimum, biasing the reported best fit solutions.  When combined, these two effects could lead to the difference in the uncertainties seen here.

These differences reflect the fundamentally distinct ways in which the two methods explore the parameter space. The GA is designed to efficiently explore the global parameter space and rapidly identify good solutions, but it does not densely sample the likelihood surface near the optimum. In contrast, MCMC is specifically designed to map the posterior distribution in detail, providing both robust parameter estimates and statistically meaningful uncertainties. Because of this, while the GA is well suited for obtaining fast initial or approximate solutions, MCMC (and NS) provide more reliable estimates of both best-fit parameters and their associated uncertainties when computational resources and time permit.

\subsection{Comparison with alternative emulator approaches}
At the time of writing, to our knowledge, three other massive star spectral emulators have been published \citep{Gull2022, Urbaneja2026, aschenbrenner2026} and several under development.  The philosophies, approaches, strengths and limitations of each differ with respect to one another, and here we attempt to summarize the major similarities and differences when compared with the approach presented in this paper.

The emulator produced by \citet{Gull2022} was created using \texttt{The Payne} \citep{Ting2019} and was trained on the TLUSTY O- and B-star grids \citep{Lanz2003, lanz2007}.  Because the TLUSTY O- and B-star grids are regularly spaced, the resulting emulator produced by \citet{Gull2022} suffers from many of the same drawbacks that a traditional grid does (see Sect. \ref{subsec:net_performance}). The major benefit of their emulator is that once trained, it requires less storage space and is less memory intensive (and therefore potentially faster) than traditional grid fitting techniques. Architecturally, both our approach and that presented in \citet{Gull2022} employ neural networks mapping stellar parameters to (normalized) fluxes, but instead of training on the entire spectral window simultaneously, we employ a line-by-line approach to the training. Furthermore, our training set is randomly sampled across the parameter space while \citet{Gull2022} uses a regularly spaced grid as a training set.  As a result, \specfann\ is less constrained by grid topology and better suited to extending parameter coverage or incorporating additional physics. The method presented in \citet{Gull2022}, on the other hand, is limited by the resolution of the regular grid over which it was trained, and therefore may struggle in poorly sampled or highly non-linear regions of the parameter space.  

The emulator produced by \citet{Urbaneja2026} was trained on \fw\ models, however their approach is fundamentally different to ours.  \citet{Urbaneja2026} uses a combination of principal component analysis \citep[PCA; ][]{connolly1995} and Gaussian process regression \citep{Rasmussen2006} to model the mapping from stellar parameters to spectra. By using PCA, the author reduces the dimensionality of the spectra, train on their principal components, and then later use these principal components to recreate the full synthetic spectrum.  On the other hand, our networks are trained directly on the line profiles without reducing the dimensionality, meaning that we are more sensitive to small scale structures and sharp transitions.  That said, the use of Gaussian process regression in the \citet{Urbaneja2026} approach provides explicit model-based uncertaity estimates, while our uncertainties come exclusively from the fitting process.  Similarly to \citet{Gull2022}, \citet{Urbaneja2026} trains on a full spectral region as opposed to individual lines as we do, meaning that incoorporating additional lines requires a full retraining of the emulator.  While our training set is sampled from a uniform distribution across the parameter space, \citet{Urbaneja2026} uses Latin-hypercube sampling, which allows for more uniform random spacing initially, but is problematic if one wants to include additional models in the training set in the future.

The emulator produced by \citet{aschenbrenner2026} is conceptually the most similar to ours of the three discussed here.  As with our approach, \citet{aschenbrenner2026} utilizes a basic feed-forward neural network structure to map stellar parameters to synthetic spectra, and draws the training set randomly across the parameter space. Our approaches diverge in two key aspects: 1) while we use \fw\ to calculate the training and test sets, \citet{aschenbrenner2026} calculates synthetic spectra from \textsc{atlas12} model atmospheres \citep{Kurucz2005} following the procedure outlined in \citet{Aschenbrenner2023}; and 2) while we train on individual lines, \citet{aschenbrenner2026} train on a spectral region similar to the previous two methods.  As mentioned in the previous comparisons, our choice to train on individual lines gives us more flexibility to include additional lines in the future and allows for a more precise mapping between the parameters and the spectral line profiles on a per-line basis.

\section{Summary and future work} \label{sec:conclusion}
In this paper we have presented a proof-of-concept neural network-based \fw\ emulator. We have detailed our \fw\ model calculations and our neural network setup, and we have shown that after training with a fairly basic network architecture, we are able to reproduce \fw\ line profiles to a high degree of accuracy, demonstrating the potential of such methods. We have also presented the \specfann\ package and detailed the implemented physics, loss functions, and fitting methods.  We applied \specfann\ to the O-type spectral standard star 10 Lac and the single O-type stars in the MELCHIORS database, and we have demonstrated that we are able to recover the stellar parameters presented in the literature in a fraction of the computation time.  Finally, we discussed the advantages that neural networks offer when compared to alternative fitting methods, and we discussed the limitations of the method as well.

In the future, we plan to improve both the neural networks and \specfann\ in several ways.  We intend to explore alternative neural network architectures to better reproduce complex wind lines, and we plan to optimize the structure of well-reproduced lines to further reduce computation time and storage space requirements.  On the \specfann\ side, we plan to implement additional cost functions and fitting techniques, including the Mahalanobis distance and Hamiltonian Monte Carlo, and we plan to implement a binary module to handle spectroscopic binary systems.  We are also working on several new bundles, some covering the same parameter space but at different metallicities and others including additional parameters such as mass loss rate, terminal wind speed, clumping, and microturbulence.  

Overall, as data from larger surveys become available, methods such as the ones presented in this paper will prove to be essential to keep up with the data flow.  \specfann\ provides the tools necessary to confront the coming era of large spectroscopic surveys in a flexible and user-friendly format, and as we have demonstrated here, it is not only able to obtain reliable stellar parameters, but it is able to do so in a fraction of the time that has been required up to this point.

\begin{acknowledgements}

This project received support from the ``La Caixa'' Foundation (ID 100010434) under the fellowship code LCF/BQ/PI23/11970035.  AE received the support of a fellowship from “La Caixa” Foundation (ID 100010434) with fellowship code LCF/BQ/PI23/11970031.

JIV and JMH acknowledge support from the European Research Council for the ERC Advanced Grant 101054731

The authors gratefully acknowledge UK Research and Innovation (UKRI) in the form of a Frontier Research grant under the UK government's ERC Horizon Europe funding guarantee (SYMPHONY; PI Bowman; grant number: EP/Y031059/1), and a Royal Society University Research Fellowship (PI Bowman; grant number: URF{\textbackslash}R1{\textbackslash}231631).

JB is supported by an NWO Veni fellowship (VI.Veni.242.199).

G.H. acknowledge support from the State Research Agency (AEI) of the Spanish Ministry of Science and Innovation (MICIN) and the European Regional Development Fund, FEDER under grants PRODUCTOS DE LA INTERACCION DE ESTRELLAS MASIVAS REVELADOS POR GRANDES SONDEOS ESPECTROSCOPICOS,  with references PID2024-159329NB-C21.

The authors wish to acknowledge the contribution of the IAC High-Performance Computing support team and hardware facilities to the results of this research. 

This paper made use of the IAC HTCondor facility (http://research.cs.wisc.edu/htcondor/), partly financed by the Ministry of Economy and Competitiveness with FEDER funds, code IACA13-3E-2493.

Based on observations made with the Mercator Telescope, operated by the Flemish Community at the Observatorio del Roque de los Muchachos (La Palma, Spain) of the Instituto de Astrofísica de Canarias. 

In addition to the codes explicitly cited in the text, this paper makes use of the following packages: \textsc{matplotlib} \citep{Hunter2007}, \textsc{numpy} \citep{Harris2020}, \textsc{scipy} \citep{Virtanen2020}, \textsc{astropy} \citep{AstropyCollaboration2013, AstropyCollaboration2018, AstropyCollaboration2022}.

\end{acknowledgements}

\bibliographystyle{aa}
\bibliography{mybib}

\onecolumn
\begin{appendix}

\section{Mean Absolute Errors per line}

\begin{table}[h!]
\caption{MAE for each spectral line in the master line list.}
\begin{minipage}{0.33\textwidth}
\begin{tabular}{lcc}
\hline
Line & MAE & $\mathrm{MAE}_\mathrm{30~km~s^{-1}}$ \\
\hline
CII1010 & 7.45e-03 & 1.63e-03 \\
CII1036 & 8.29e-03 & 3.00e-03 \\
CII1066 & 4.04e-03 & 1.12e-03 \\
CII1324 & 1.84e-03 & 5.64e-04 \\
CII1335 & 5.44e-03 & 2.38e-03 \\
CII1760 & 1.71e-03 & 5.82e-04 \\
CII3920 & 5.60e-04 & 2.93e-04 \\
CII4267 & 9.38e-04 & 3.10e-04 \\
CII5133 & 2.31e-04 & 1.19e-04 \\
CII5137 & 8.54e-05 & 5.97e-05 \\
CII5139 & 6.32e-05 & 2.78e-05 \\
CII5143 & 1.15e-04 & 5.68e-05 \\
CII5145 & 1.86e-04 & 1.14e-04 \\
CII5151 & 1.03e-04 & 4.11e-05 \\
CII5333 & 6.44e-05 & 2.84e-05 \\
CII5335 & 6.93e-05 & 3.53e-05 \\
CII5641 & 1.06e-04 & 7.54e-05 \\
CII5648 & 8.08e-05 & 3.17e-05 \\
CII5662 & 1.19e-04 & 7.07e-05 \\
CII5890 & 3.39e-04 & 1.93e-04 \\
CII6151 & 2.83e-04 & 1.98e-04 \\
CII6462 & 2.16e-04 & 1.43e-04 \\
CII6527 & 7.08e-05 & 4.19e-05 \\
CII6530 & 8.26e-05 & 3.51e-05 \\
CII6578 & 5.46e-04 & 2.37e-04 \\
CII6582 & 4.66e-04 & 2.36e-04 \\
CII6780 & 1.59e-04 & 7.76e-05 \\
CII6784 & 2.32e-04 & 1.62e-04 \\
CII6787 & 7.70e-05 & 3.57e-05 \\
CII6791 & 1.06e-04 & 5.84e-05 \\
CII6798 & 6.27e-05 & 3.63e-05 \\
CII6801 & 8.40e-05 & 4.09e-05 \\
CII6812 & 5.64e-05 & 3.67e-05 \\
CII7231 & 2.98e-04 & 1.27e-04 \\
CII7237 & 3.92e-04 & 1.29e-04 \\
CIII1176 & 1.71e-02 & 6.90e-03 \\
CIII1247 & 3.00e-03 & 1.39e-03 \\
CIII1620 & 2.04e-03 & 6.44e-04 \\
CIII4056 & 4.54e-04 & 3.11e-04 \\
CIII4070 & 4.56e-04 & 1.59e-04 \\
CIII4162 & 3.68e-04 & 2.67e-04 \\
CIII4186 & 4.64e-04 & 1.79e-04 \\
CIII4650 & 1.84e-03 & 5.80e-04 \\
CIII4665 & 3.95e-04 & 2.66e-04 \\
CIII5696 & 8.09e-04 & 2.97e-04 \\
CIII5826 & 3.99e-04 & 2.13e-04 \\
CIII6744 & 2.46e-04 & 1.89e-04 \\
\hline
\end{tabular}
\end{minipage}
\hfill
\begin{minipage}{0.33\textwidth}
\begin{tabular}{lcc}
\hline
Line & MAE & $\mathrm{MAE}_\mathrm{30~km~s^{-1}}$ \\
\hline
CIII8500 & 4.04e-04 & 1.76e-04 \\
CIV1169 & 2.88e-03 & 6.49e-04 \\
CIV1550 & 2.09e-02 & 1.36e-02 \\
CIV5018 & 1.60e-04 & 1.08e-04 \\
CIV5801 & 3.32e-03 & 1.61e-03 \\
CIV5811 & 1.43e-03 & 6.13e-04 \\
HALPHA & 7.72e-03 & 2.66e-03 \\
HBETA & 2.03e-03 & 8.57e-04 \\
HGAMMA & 1.49e-03 & 6.55e-04 \\
HDELTA & 2.12e-03 & 8.32e-04 \\
HEPS & 1.07e-03 & 4.89e-04 \\
HEI4009 & 6.14e-04 & 2.51e-04 \\
HEI4026 & 1.43e-03 & 5.69e-04 \\
HEI4121 & 1.43e-03 & 5.23e-04 \\
HEI4143 & 8.52e-04 & 3.38e-04 \\
HEI4387 & 7.75e-04 & 3.28e-04 \\
HEI4471 & 9.79e-04 & 4.21e-04 \\
HEI4713 & 6.56e-04 & 2.52e-04 \\
HEI4922 & 9.99e-04 & 3.80e-04 \\
HEI5016 & 7.36e-04 & 2.74e-04 \\
HEI5048 & 7.01e-04 & 3.17e-04 \\
HEI5875 & 1.19e-03 & 5.01e-04 \\
HEI6678 & 1.50e-03 & 5.53e-04 \\
HEI7065 & 9.60e-04 & 3.15e-04 \\
HEII1640 & 3.43e-03 & 2.18e-03 \\
HEII4200 & 1.06e-03 & 4.75e-04 \\
HEII4541 & 5.50e-04 & 3.44e-04 \\
HEII4686 & 8.97e-03 & 8.24e-03 \\
HEII5411 & 9.58e-04 & 7.39e-04 \\
HEII6406 & 3.38e-04 & 2.28e-04 \\
HEII6527 & 4.85e-04 & 3.58e-04 \\
HEII6683 & 1.53e-03 & 6.22e-04 \\
NII3995 & 6.74e-04 & 2.75e-04 \\
NII4041 & 7.32e-04 & 2.67e-04 \\
NII4447 & 3.46e-04 & 1.41e-04 \\
NII4530 & 2.75e-04 & 1.02e-04 \\
NII4552 & 2.47e-04 & 1.51e-04 \\
NII4601 & 3.81e-04 & 1.75e-04 \\
NII4607 & 2.52e-04 & 1.07e-04 \\
NII4621 & 4.48e-04 & 2.75e-04 \\
NII4630 & 5.37e-04 & 2.29e-04 \\
NII4643 & 4.59e-04 & 2.19e-04 \\
NII5005 & 5.03e-04 & 2.09e-04 \\
NII5007 & 6.04e-04 & 3.72e-04 \\
NII5045 & 4.70e-04 & 2.55e-04 \\
NII5666 & 3.51e-04 & 1.49e-04 \\
NII5676 & 2.49e-04 & 9.69e-05 \\
\hline
\end{tabular}
\end{minipage}
\hfill
\begin{minipage}{0.33\textwidth}
\begin{tabular}{lcc}
\hline
Line & MAE & $\mathrm{MAE}_\mathrm{30~km~s^{-1}}$ \\
\hline
NII5679 & 5.18e-04 & 2.05e-04 \\
NII5710 & 3.49e-04 & 2.25e-04 \\
NII5931 & 1.97e-04 & 8.61e-05 \\
NII5941 & 2.65e-04 & 9.91e-05 \\
NII6482 & 7.62e-04 & 3.03e-04 \\
NII6610 & 2.98e-04 & 1.14e-04 \\
NIII3355 & 4.99e-04 & 1.90e-04 \\
NIII3366 & 7.22e-04 & 3.71e-04 \\
NIII3999 & 3.80e-04 & 2.25e-04 \\
NIII4003 & 4.02e-04 & 1.44e-04 \\
NIII4097 & 9.85e-04 & 3.71e-04 \\
NIII4103 & 7.11e-04 & 2.79e-04 \\
NIII4379 & 7.37e-04 & 2.15e-04 \\
NIII4523 & 4.08e-04 & 2.14e-04 \\
NIII4527 & 1.77e-04 & 1.18e-04 \\
NIII4531 & 1.21e-04 & 5.74e-05 \\
NIII4535 & 7.47e-04 & 3.89e-04 \\
NIII4547 & 2.49e-04 & 9.99e-05 \\
NIIIqua & 9.03e-04 & 2.81e-04 \\
NIIItrip & 1.70e-03 & 5.47e-04 \\
NIV1718 & 3.06e-03 & 2.05e-03 \\
NIV3480 & 1.86e-03 & 8.33e-04 \\
NIV4058 & 1.18e-03 & 3.74e-04 \\
NIV5205 & 6.53e-04 & 3.13e-04 \\
NIV6380 & 7.84e-04 & 3.52e-04 \\
NV1240 & 1.12e-02 & 7.63e-03 \\
NVall & 5.20e-04 & 2.97e-04 \\
OII3954 & 2.58e-04 & 1.13e-04 \\
OII4072 & 4.06e-04 & 1.73e-04 \\
OII4075 & 7.93e-04 & 3.51e-04 \\
OII441416 & 6.09e-04 & 2.53e-04 \\
OII4661 & 4.13e-04 & 2.58e-04 \\
OII4906 & 1.65e-04 & 6.07e-05 \\
OII4941 & 3.94e-04 & 2.45e-04 \\
OIII3961 & 6.48e-04 & 2.12e-04 \\
OIII4081 & 2.36e-04 & 1.44e-04 \\
OIII5268 & 4.04e-04 & 1.95e-04 \\
OIII5508 & 1.77e-04 & 7.01e-05 \\
OIII5592 & 4.83e-04 & 1.45e-04 \\
OIV1340 & 6.78e-03 & 4.58e-03 \\
OIV4654 & 3.11e-04 & 1.61e-04 \\
OV1371 & 3.52e-03 & 2.47e-03 \\
SIIV4631 & 4.62e-04 & 1.94e-04 \\
SIIV4654 & 7.95e-04 & 2.88e-04 \\
SIIV6667 & 4.29e-04 & 1.81e-04 \\
SIIV6701 & 8.97e-04 & 2.76e-04 \\
 & & \\
\hline
\end{tabular}
\end{minipage}
\tablefoot{
The first column indicates the name of the line, the second column indicates the raw global MAE and the third column indicates the global MAE after convolving the line profiles with a 30 \kms rotation kernel.  In most cases, the line name indicates the element, ionization stage and central wavelength (in $\AA$).  Exceptions include the Balmer lines, two \nc\ multiplets, an \ne\ multiplet and an \ob\ doublet.  The central wavelengths (in $\AA$) for the Balmer lines are: HALPHA - 6563, HBETA - 4861, HGAMMA - 4340, HDELTA - 4102, HEPS - 3970, and the range of central values of the multiplets are NIIIqua $\sim$ 4511-4518, NIIItrip $\sim$ 4634-4641, NVall $\sim$ 4603-4619, OII441416 $\sim$ 4414-4416.}
\label{tab: MAEs}
\end{table}

\newpage
\section{Validation plots}

\begin{figure}[h!]
    \centering
    \includegraphics[width=0.4\textwidth]{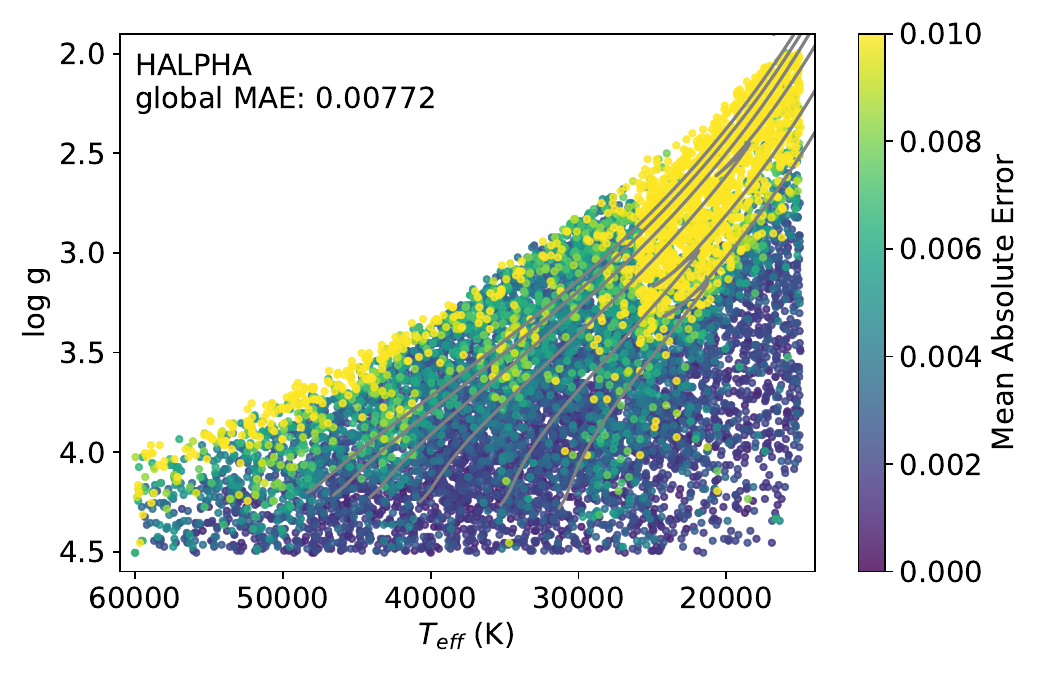}
    \includegraphics[width=0.4\textwidth]{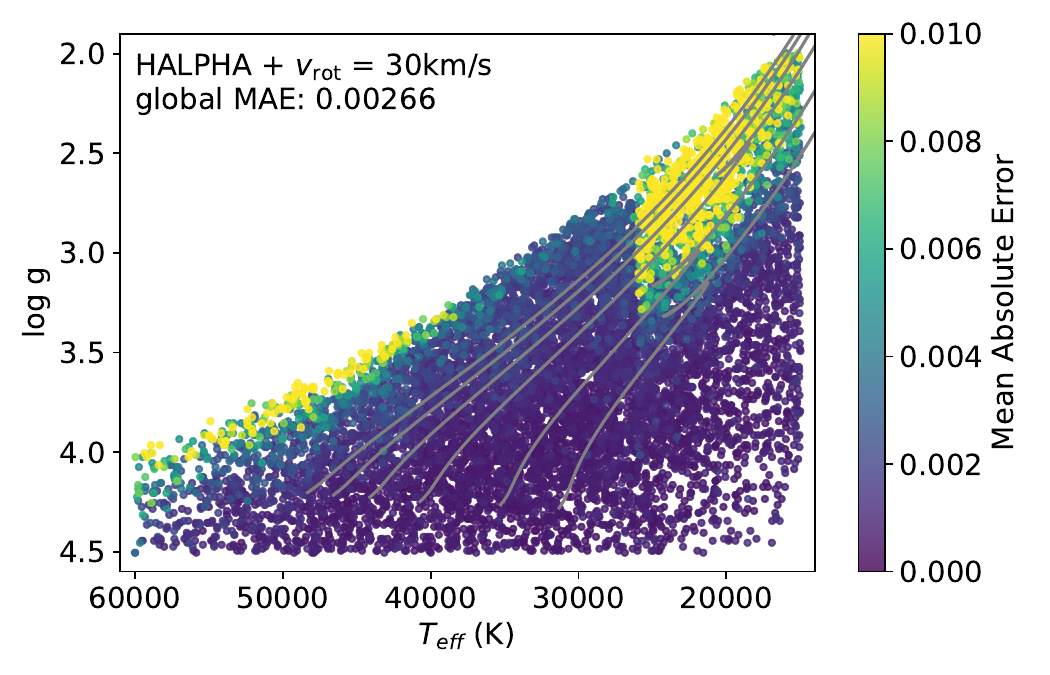}

    \includegraphics[width = 0.8\textwidth]{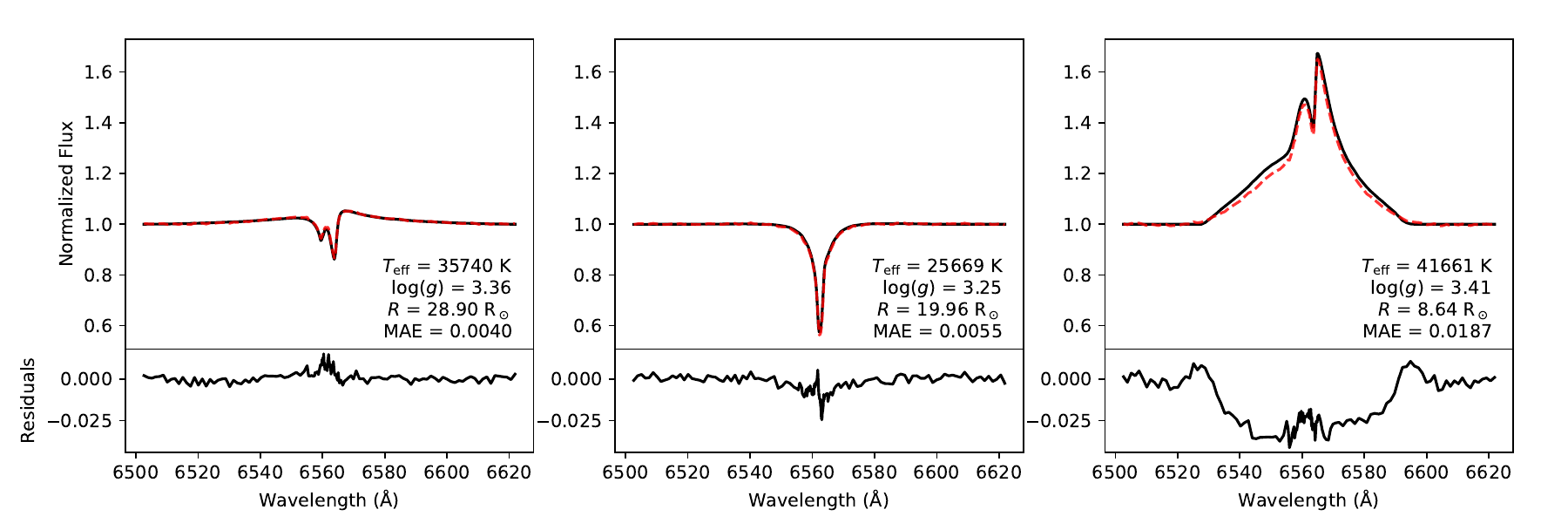}

    \caption{Same as Fig. \ref{fig: HeI4471_validation}, but for the \halpha\ line.}
    \label{fig: HALPHA_validation}
\end{figure}

\begin{figure}
    \centering
    \includegraphics[width=0.4\textwidth]{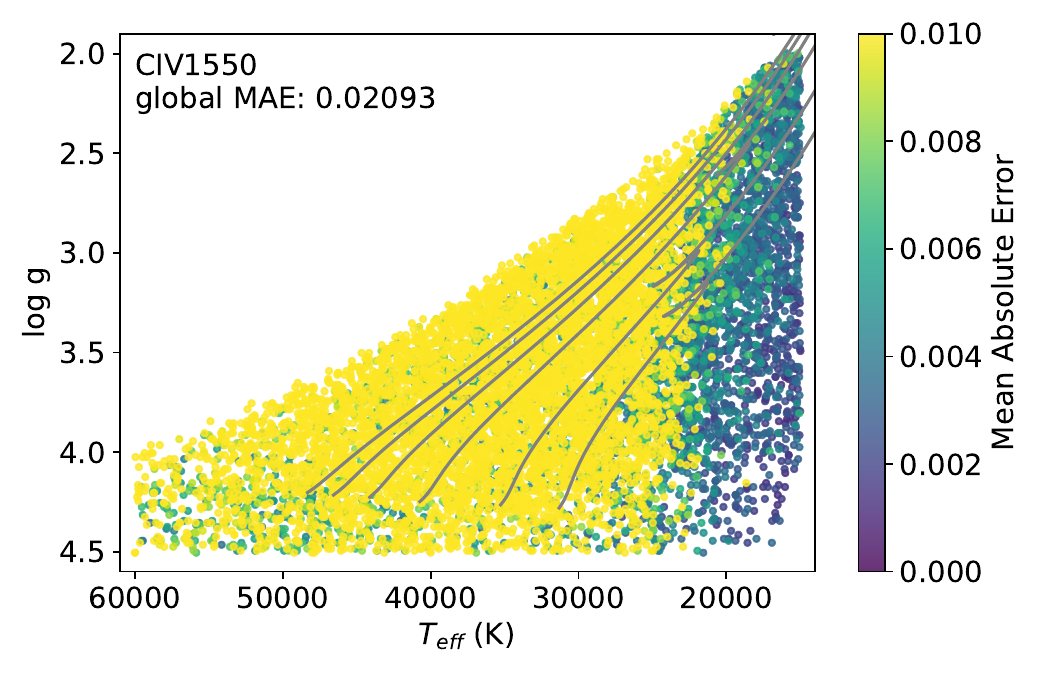}
    \includegraphics[width=0.4\textwidth]{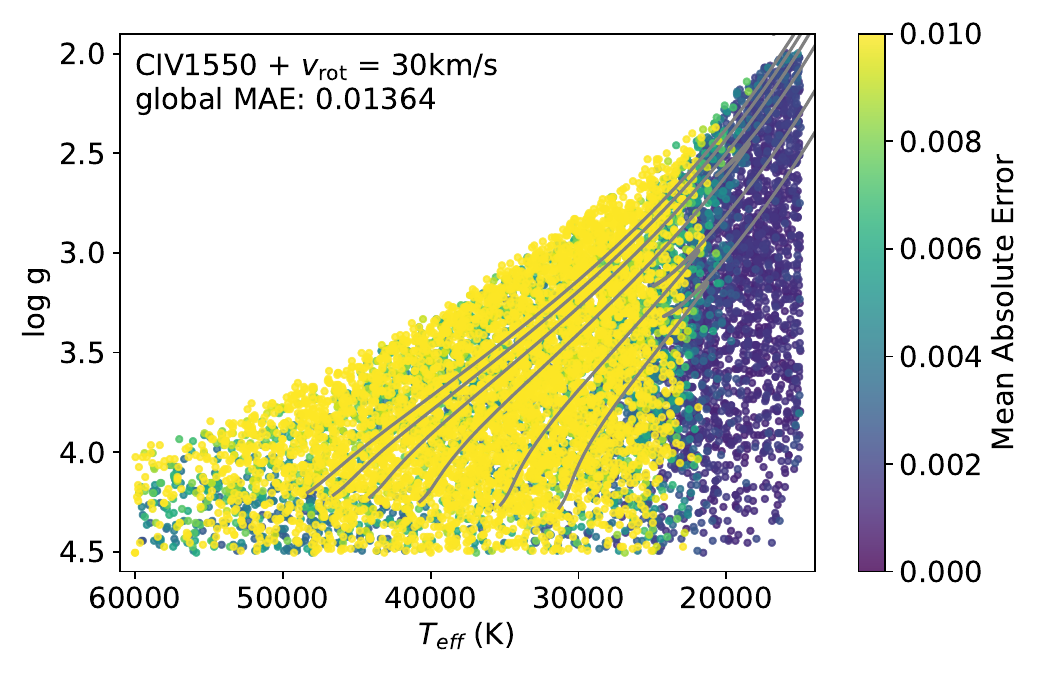}

    \includegraphics[width = 0.8\textwidth]{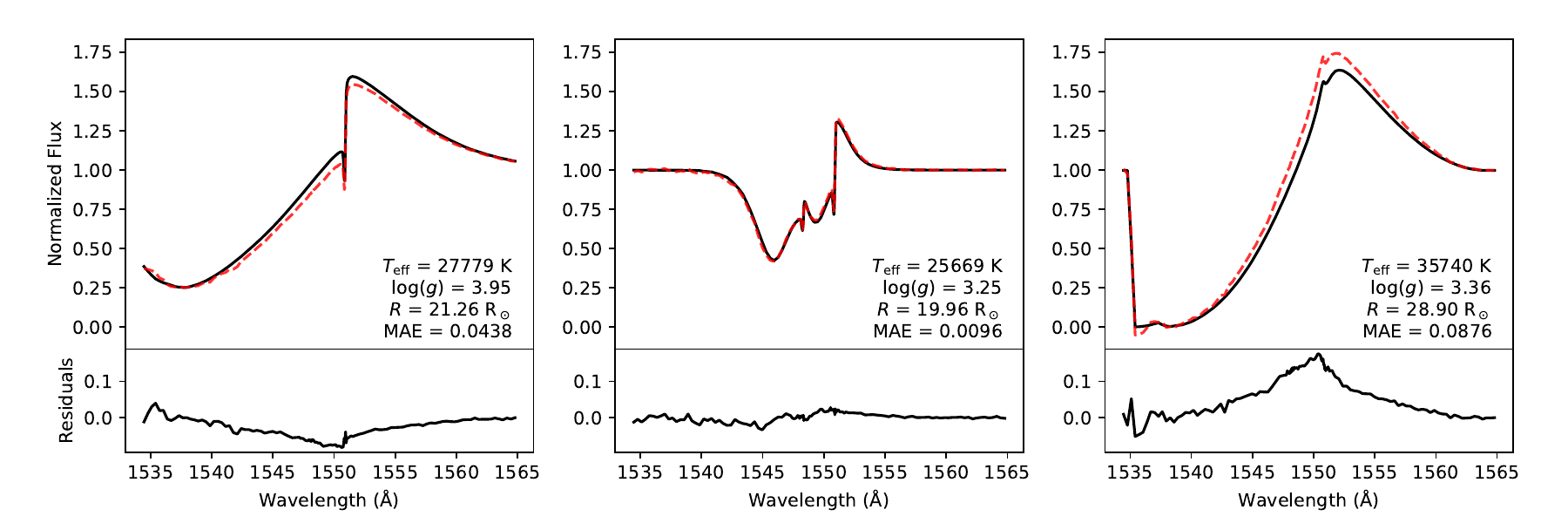}

    \caption{Same as Fig. \ref{fig: HeI4471_validation}, but for the \cd \lam1550 line.}
    \label{fig: CIV1550_validation}
\end{figure}

\end{appendix}
\twocolumn
\end{document}